# Extensions of the Novikov-Furutsu theorem, obtained by using Volterra functional calculus

G.A. Athanassoulis and K.I. Mamis
mathan@central.ntua.gr, konmamis@central.ntua.gr

School of Naval Architecture & Marine Engineering, National Technical University of Athens,
9 Iroon Polytechniou st., 15780 Zografos, GREECE

**Abstract.** Novikov-Furutsu (NF) theorem is a well-known mathematical tool, used in stochastic dynamics for correlation splitting, that is, for evaluating the mean value of the product of a random functional with a Gaussian argument multiplied by the argument itself. In this work, the NF theorem is extended for mappings (function-functionals) of two arguments, one being a random variable and the other a random function, both of which are Gaussian, may have non-zero mean values, and may be correlated with each other. This extension allows for the study of random differential equations under coloured noise excitation, which may be correlated with the random initial value. Applications in this direction are briefly discussed. The proof of the extended NF theorem is based on a more general result, also proven herein by using Volterra functional calculus, stating that: The mean value of a general, nonlinear function-functional having random arguments, possibly non-Gaussian, can be expressed in terms of the characteristic functional of its arguments. Generalizations to the multidimensional case (multivariate random arguments) are also presented.

**TABLE OF CONTENTS**

## 1. Introduction

In their works (Furutsu, 1963; Novikov, 1965), on electromagnetic waves and turbulence respectively, K. Furutsu and E. Novikov derived independently the following formula ([1])

$$\mathbb{E}^{\theta}\left[\Xi(t;\theta)\,\mathcal{F}[\Xi(\bullet|_{t_0}^{t};\theta)]\right]=\int_{t_0}^{t}C_{\Xi(\bullet)\Xi(\bullet)}(t,s)\,\mathbb{E}^{\theta}\left[\frac{\delta\mathcal{F}[\Xi(\bullet|_{t_0}^{t};\theta)]}{\delta\Xi(s;\theta)}\right]ds,$$

where $\Xi(\bullet;\theta)$ is a zero-mean Gaussian random function with autocovariance function $C_{\Xi(\bullet)\Xi(\bullet)}(t,s)$, $\mathcal{F}[\Xi(\bullet|_{t_0}^{t};\theta)]$ is a functional of $\Xi(\bullet;\theta)$ over $[t_0,t]$, $\mathbb{E}^{\theta}[\bullet]$ denotes the mean value operator, and $\delta/\delta\Xi(s;\theta)$ is the Volterra functional derivative with respect to $\Xi(\bullet;\theta)$ at $s$. The above equation, called henceforth the Novikov-Furutsu (NF) theorem, has been extensively used in *turbulent diffusion* (Cook, 1978; Hyland, McKee and Reeks, 1999; Shrimpton, Haeri and Scott, 2014; Krommes, 2015; Martins Afonso, Mazzino and Gama, 2016), *random waves* (Sobczyk, 1985; Rino, 1991; Bobryk, 1993; Konotop and Vazquez, 1994; Creamer, 2008), and *stochastic dynamics* (Protopopescu, 1983; Klyatskin, 2005, 2015; Scott, 2013; Kliemann and Sri Namachchivaya, 2018). One particular use of the NF theorem lies in the formulation of partial differential equations, called generalized Fokker-Planck-Kolmogorov (genFPK) equations, that govern the evolution of the probability density function (pdf) of the response to a random differential equation (RDE) excited by coloured noise (Sancho *et al.*, 1982; Cetto and de la Peña, 1984; Fox, 1986; Hyland, McKee and Reeks, 1999; Venturi *et al.*, 2012).

In this work, the classical NF theorem is extended (Theorem 2 in Section 2) to random function-functionals ($\mathrm{RFF}\ell$) $\mathcal{F}[X_0(\theta);\Xi(\bullet|_{t_0}^{t};\theta)]$ (function on the scalar random variable $X_0(\theta)$ and functional on the scalar random function $\Xi(\bullet;\theta)$ over the interval $[t_0,t]$). $\mathrm{RFF}\ell$'s arguments $X_0(\theta)$, $\Xi(\bullet;\theta)$ are considered correlated and jointly Gaussian, with mean values $m_{X_0}$, $m_{\Xi(\bullet)}(\bullet)$, autocovariances, $C_{X_0X_0}$, $C_{\Xi(\bullet)\Xi(\bullet)}(\bullet,\bullet)$, and cross-covariance $C_{X_0\Xi(\bullet)}(\bullet)$. This extension of the NF theorem is useful for the formulation of pdf equations corresponding to RDEs in which both the initial value and the excitation are Gaussian random elements, correlated to each other; see Section 6.

What is more, the above extension of the NF theorem is derived as a special case of a more general result (Theorem 1 in Section 2), whose importance has not been highlighted before, to the best of our knowledge. Theorem 1 expresses the mean value of a general, nonlinear $\mathrm{RFF}\ell$ in terms of the joint characteristic function-functional ($\mathrm{j.Ch.FF}\ell$) of its arguments, which can be non-Gaussian.

Theorems 1, 2 are proved in Sections 4 and 5 respectively, by using the mathematical tools introduced by Vito Volterra, namely the concept of passing from the discrete to continuous and his eponymous calculus for functionals, which are presented briefly in Appendix A. An application of the extended NF Theorem to the derivation of an evolution equation governing the response probability density function of ordinary differential equations under coloured excitation is briefly presented in Section 6. In Section 7, we formulate and study the exact response pdf evolution equation for a linear RDE under Gaussian coloured noise correlated to its Gaussian initial value. The unique solution of this equation has been found, and it turns to be identical with the correct Gaussian response pdf (obtained through the corresponding linear moment problem), as expected. Since the derivation of the said evolution equation is based on the extended NF Theorem 2, this example also constitutes a first validation for Theorem 2. In Section 8, we state (without complete proofs) the generalizations of Theorems 1 and 2 to the multidimensional case, i.e. when $\mathbf{X}_0(\theta)$ and $\Xi(\bullet;\theta)$ are a random vector and a random vector function, respectively. In the concluding Section 9, a recapitulation of the results is given, while the merits of the novel proof of NF theorem presented herein and further generalizations, are discussed.

([1]) $\theta$ denotes the stochastic argument.

## 2. The mean value of a random, nonlinear function-functional

Before proceeding with the study of the $\mathrm{RFF}\ell$, we shall introduce its deterministic counterpart and state the basic analytical assumptions. Consider a real-valued function-functional, $\mathcal{G}[\upsilon\,;u(\bullet\big|_{t_0}^{t})]:\mathbb{R}\times\mathscr{Z}\to\mathbb{R}$, where $\mathscr{Z}=C\big([t_0,t]\to\mathbb{R}\big)$ is the space of continuous functions. The function-functional $\mathcal{G}[\upsilon\,;u(\bullet\big|_{t_0}^{t})]$ is assumed to have derivatives of any order, both with respect to the scalar argument $\upsilon$ and the function argument $u(\bullet)$, and it is expandable in Volterra-Taylor series, jointly with respect to $\upsilon$ and $u(\bullet)$, around a fixed pair $\big(\upsilon_0\,;u_0(\bullet)\big)$. Here, functional derivatives are considered in the sense of Volterra (Volterra, 1930), which are reconsidered in a rigorous manner by (Donsker and Lions, 1962). For the sake of brevity, the above smoothness conditions of $\mathcal{G}[\upsilon\,;u(\bullet\big|_{t_0}^{t})]$ will be denoted as $C^\infty$ henceforth.

The $\mathrm{RFF}\ell$ is obtained by replacing the argument $\big(\upsilon\,;u(\bullet)\big)$ by the random element $\big(X_0(\theta)\,;\Xi(\bullet;\theta)\big)$. The latter is fully described by the infinite-dimensional joint probability measure $\mathbf{P}_{X_0\,\Xi(\bullet)}$, defined over the Borel $\sigma-$algebra $\mathscr{B}(\mathbb{R}\times\mathscr{Z})$. Probability measure $\mathbf{P}_{X_0\,\Xi(\bullet)}$ is assumed to be continuously distributed, i.e. it possesses a well-defined probability density functional. Since $\big(X_0(\theta)\,;\Xi(\bullet;\theta)\big):\theta\to\mathbb{R}\times\mathscr{Z}$ is Borel measurable, and $\mathcal{G}[\upsilon\,;u(\bullet\big|_{t_0}^{t})]$ is $C^\infty$, the composition $\mathcal{G}[X_0(\theta)\,;\Xi(\bullet\big|_{t_0}^{t};\theta)]$ is also $\mathscr{B}(\mathbb{R}\times\mathscr{Z})-$measurable.

Then, by definition, the mean value of $\mathcal{G}[X_0(\theta)\,;\Xi(\bullet\big|_{t_0}^{t};\theta)]$ is expressed as

$$\mathbb{E}^{\theta}_{\mathbf{P}_{X_0\,\Xi(\bullet)}}\Big[\mathcal{G}[X_0(\theta)\,;\Xi(\bullet\big|_{t_0}^{t};\theta)]\Big]=\int_{\mathbb{R}\times\mathscr{Z}}\mathcal{G}[\chi\,;\xi(\bullet)]\,\mathbf{P}_{X_0\,\Xi(\bullet)}\big(d\chi\times d\xi(\bullet)\big). \tag{1}$$

However, the above definition involves an integral over an infinite-dimensional space, which, in most of the cases, is rather difficult to calculate. As a more easily computable alternative, we express, in the following theorem, the mean value of a $\mathrm{RFF}\ell$, via the probabilistic structure of its arguments, as described by its joint characteristic function-functional ($\mathrm{j.Ch.FF}\ell$):

$$\varphi_{X_0\,\Xi(\cdot)}[\upsilon\,;u(\bullet|_{t_0}^{t})] = \mathbb{E}^{\theta}_{\mathbf{P}_{X_0\,\Xi(\cdot)}}\left[\exp\left(i\,X_0(\theta)\,\upsilon + i\int_{t_0}^{t}\Xi(s\,;\theta)\,u(s)\,ds\right)\right]. \tag{2}$$

Note that, the random elements $X_0(\theta)$, $\Xi(\bullet\,;\theta)$ may be dependent, having *any prescribed probability distribution*.

**Theorem 1 [Mean value of an** RFF$\ell$**]:** The mean value of the random function-functional $\mathcal{G}[X_0(\theta)\,;\Xi(\bullet|_{t_0}^{t}\,;\theta)]$ is expressed by the formula

$$\mathbb{E}^{\theta}\Big[\mathcal{G}[X_0(\theta)\,;\Xi(\bullet|_{t_0}^{t}\,;\theta)]\Big] = \varphi_{\hat{X}_0\,\hat{\Xi}(\cdot)}\left(\frac{\partial}{i\,\partial\upsilon}\,;\frac{\delta}{i\,\delta u(\cdot)}\right)\mathcal{G}[\upsilon;u(\bullet|_{t_0}^{t})]\Big|_{\substack{\upsilon=m_{X_0}\\ u(\bullet)=m_{\Xi(\cdot)}(\bullet)}} =$$

$$= \mathbb{E}^{\theta}\left[\exp\left(\hat{X}_0(\theta)\,\frac{\partial}{\partial\upsilon} + \int_{t_0}^{t}ds\,\hat{\Xi}(s\,;\theta)\,\frac{\delta}{\delta u(s)}\right)\right]\mathcal{G}[\upsilon;u(\bullet|_{t_0}^{t})]\Big|_{\substack{\upsilon=m_{X_0}\\ u(\bullet)=m_{\Xi(\cdot)}(\bullet)}} \tag{3}$$

where $\mathbb{E}^{\theta}[\,\bullet\,]\equiv\mathbb{E}^{\theta}_{\mathbf{P}_{X_0\,\Xi(\cdot)}}[\,\bullet\,]$, $m_{X_0}$ and $m_{\Xi(\cdot)}(s)$ are the mean values of $X_0(\theta)$ and $\Xi(\bullet\,;\theta)$, respectively, $\partial/\partial\upsilon$ denotes partial differentiation with respect to $\upsilon$, and $\delta/\delta u(s)$ denotes Volterra functional differentiation with respect to the function $u(\bullet)$ at $s$. Further, the quantities

$$\hat{X}_0(\theta) = X_0(\theta) - m_{X_0}, \qquad \hat{\Xi}(s;\theta) = \Xi(s;\theta) - m_{\Xi(\cdot)}(s) \tag{4a,b}$$

are the fluctuations of the random elements $X_0(\theta)$ and $\Xi(s;\theta)$ around their mean values, and $\varphi_{\hat{X}_0\,\hat{\Xi}(\cdot)}[\upsilon\,;u(\bullet)]$ is the j.Ch.FF$\ell$ of the said fluctuations. The exact meaning of the operator appearing in the right-hand side of Eq. (3) will be defined below, during the derivation of Eq. (3). ■

**Remark 1.** By comparing formula (3) to Eq. (1), we observe that, by Theorem 1, we compute an integration over an infinite-dimensional space by the action of a pseudo-differential operator. A similar practice of integration by differentiation in functional spaces, albeit with quite different applications, has been recently introduced by Kempf, Jackson and Morales (2014, 2015); Jia, Tang and Kempf (2017) in quantum field theory.

**Remark 2.** A formula similar to Eq. (3) for a random functional of the form $\mathcal{G}[\Xi(\,\bullet\,;\theta)] = \mathcal{G}_1[\Xi(\,\bullet\,;\theta)]\,\mathcal{G}_2[\Xi(\,\bullet\,;\theta)]$ is given by (Klyatskin, 2005), Ch. 4. The proof given by Klyatskin is concise and unmotivated, thus difficult to follow for a reader not well-acquainted with Volterra functional calculus. Eq. (3) includes Klyatskin's result as a special case. The proof presented herein is fairly detailed and motivated by easily understood analogues in the discrete setting. Then, the Volterra technique of passing from the discrete to continuous reveals the appropriate forms in the function-functional setting. The latter can be directly proved by using Volterra functional calculus.

The proof of Theorem 1 is completed in Section 4, and is based on the expansion of the RFF$\ell$ $\mathcal{G}[X_0(\theta);\Xi(\bullet|_{t_0}^{t};\theta)]$ around the mean values $m_{X_0}$ and $m_{\Xi}(\bullet)$, by employing the function-functional shift operator. The latter is introduced in Section 3, via Volterra's concept of passing from the discrete to continuous. This principle, discussed concisely in Appendix A, motivates a calculus for functionals that is analogous to the calculus for functions of many variables, which can be (and has been) rigorous; see also Appendix A for relevant discussion and references.

Now, by setting in Eq. (3):

$$\begin{aligned}&\mathcal{G}[X_0(\theta);\Xi(\bullet|_{t_0}^{t};\theta)] \equiv \mathcal{G}[\cdots] = \\ &\qquad = \Xi(t;\theta)\,\mathcal{F}[X_0(\theta);\Xi(\bullet|_{t_0}^{t};\theta)] \equiv \Xi(t;\theta)\,\mathcal{F}[\cdots],\end{aligned} \tag{5}$$

specifying the j.Ch.FF$\ell$ $\varphi_{X_0\,\Xi(\bullet)}[\upsilon;u(\bullet)]$ as a Gaussian one (see Appendix A for its derivation):

$$\begin{aligned}&\varphi^{\text{Gauss}}_{X_0\,\Xi(\bullet)}[\upsilon;u(\bullet|_{t_0}^{t})] = \\ &\quad = \exp\left( i\int_{t_0}^{t} m_{\Xi(\bullet)}(s)u(s)\,ds - \frac{1}{2}\int_{t_0}^{t}\int_{t_0}^{t} C_{\Xi(\bullet)\Xi(\bullet)}(s_1,s_2)\,u(s_1)\,u(s_2)\,ds_1 ds_2 \right)\times \\ &\qquad\qquad \times \exp\left( i m_{X_0}\,\upsilon - \frac{1}{2}C_{X_0X_0}\,\upsilon^2\right)\cdot\exp\left(-\upsilon\int_{t_0}^{t} C_{X_0\Xi(\bullet)}(s)\,u(s)\,ds\right),\end{aligned} \tag{6}$$

and calculating the action of the operator $\varphi_{\hat{X}_0\,\hat{\Xi}(\bullet)}\left(\dfrac{\partial}{i\,\partial\upsilon};\dfrac{\delta}{i\,\delta u(\bullet)}\right)$ on the function-functional $u(t)\,\mathcal{F}[\upsilon;u(\bullet|_{t_0}^{t})]$, the following extension of the NF theorem for RFF$\ell$s is obtained.

**Theorem 2 [The extended Novikov-Furutsu theorem]:** For a sufficiently smooth RFF$\ell$ of the form $\mathcal{F}[X_0(\theta);\Xi(\bullet|_{t_0}^{t};\theta)] \equiv \mathcal{F}[\cdots]$, whose arguments $X_0(\theta)$, $\Xi(\bullet;\theta)$ are jointly Gaussian, the following formula holds true

$$\begin{aligned}&\mathbb{E}^{\theta}\big[\Xi(t;\theta)\,\mathcal{F}[\cdots]\big] = \\ &\quad = m_{\Xi(\bullet)}(t)\,\mathbb{E}^{\theta}\big[\mathcal{F}[\cdots]\big] + C_{X_0\Xi(\bullet)}(t)\,\mathbb{E}^{\theta}\left[\frac{\partial\mathcal{F}[\cdots]}{\partial X_0(\theta)}\right] + \\ &\qquad\qquad + \int_{t_0}^{t} C_{\Xi(\bullet)\Xi(\bullet)}(t,s)\,\mathbb{E}^{\theta}\left[\frac{\delta\mathcal{F}[\cdots]}{\delta\Xi(s;\theta)}\right]ds. \qquad \blacksquare\end{aligned} \tag{7}$$

Detailed proof of Theorem 2 is given in Section 5. Note that the extension of the classical NF theorem for a non-zero mean $\Xi(t;\theta)$ uncorrelated to $X_0(\theta)$, that gives rise to the term $m_{\Xi(\bullet)}(t)\, \mathbb{E}^{\theta}\left[\mathcal{F}[\cdots]\right]$ in Eq. (7), has also been given in (Hänggi, 1989; Cáceres, 2017, Sec. 1.12.2). The general form of Eq. (7), corresponding to correlated $X_0(\theta)$ and $\Xi(t;\theta)$, is novel, to the best of our knowledge.

## 3. The function-functional shift operator and its exponential form

The shift operator for function-functionals is constructed by using Volterra's passing from the discrete to continuous, as explained in Appendix A. As a first step, we introduce its discrete analogue, the multivariate shift operator $\mathcal{T}_{\hat{\mathbf{x}}}$ for functions of many variables, $g(\mathbf{x}_0+\hat{\mathbf{x}}) = \mathcal{T}_{\hat{\mathbf{x}}}\, g(\mathbf{x}_0)$, which is in fact a compact form of the Taylor series expansion.

Let $g(\bullet) \in C^{\infty}(\mathbb{R}^{N+1} \to \mathbb{R})$. Then, the Taylor expansion of $g(\mathbf{x}) = g(\mathbf{x}_0+\hat{\mathbf{x}})$ around the point $\mathbf{x}_0 \in \mathbb{R}^{N+1}$, reads as

$$g(\mathbf{x}_0+\hat{\mathbf{x}}) = \left(1+\sum_{m=1}^{\infty}\frac{1}{m!}\sum_{n_1=1}^{N+1}\overset{(m)}{\cdots}\sum_{n_m=1}^{N+1}\hat{x}_{n_1}\cdots\hat{x}_{n_m}\frac{\partial^m}{\partial x_{n_1}\cdots\partial x_{n_m}}\right) g(\mathbf{x}_0). \tag{8}$$

To write the above equation in a more concise form, the following identity is used

$$\left(\sum_{n=1}^{N+1} h_n\, u_n\right)^m \equiv \prod_{i=1}^{m}\sum_{n_i=1}^{N+1} h_{n_i}\, u_{n_i} = \sum_{n_1=1}^{N+1}\overset{(m)}{\cdots}\sum_{n_m=1}^{N+1} h_{n_1}\cdots h_{n_m}\cdot u_{n_1}\cdots u_{n_m}, \tag{9}$$

which is valid under the assumption that all symbols $h_{n_i}, u_{n_i}$ commute, or their order is definitely prescribed, as dictated by the specific case. Setting $h_n = \hat{x}_n$ and $u_n = \partial\bullet/\partial x_n$, and imposing the order convention that all $\hat{x}_n$ precede of all $\partial\bullet/\partial x_n$, we can apply Eq. (9) to express the differential operator appearing in the right-hand side of Eq. (8), recasting the latter in the following compact form:

$$g(\mathbf{x}_0+\hat{\mathbf{x}}) = \left[\sum_{m=0}^{\infty}\frac{1}{m!}\left(\sum_{n=1}^{N+1}\hat{x}_n\frac{\partial}{\partial x_n}\right)^m\right] g(\mathbf{x}_0). \tag{10}$$

In Eq. (10) the additional (trivial) convention $\left(\sum_{n=1}^{N+1}\hat{x}_n\,\partial\bullet/\partial x_n\right)^0 = 1$ is used. Then, we observe that the infinite series appearing in the right-hand side of Eq. (10), under the said order assumption, is identified as the Taylor series of the symbol $\exp\left(\sum_{n=1}^{N+1}\hat{x}_n\,\partial\bullet/\partial x_n\right)$, leading to the nice formula

$$g(\mathbf{x}_0+\hat{\mathbf{x}}) = \exp\left(\sum_{n=1}^{N+1}\hat{x}_n\frac{\partial}{\partial x_n}\right) g(\mathbf{x}_0) = \exp\left(\hat{\mathbf{x}}\cdot\nabla_{\mathbf{x}}\right) g(\mathbf{x}_0). \tag{11}$$

Comparing Eq. (11) with $g(\mathbf{x}_0+\hat{\mathbf{x}}) = \mathcal{T}_{\hat{\mathbf{x}}}\, g(\mathbf{x}_0)$, we see that the symbol $\exp\left(\hat{\mathbf{x}}\cdot\nabla_{\mathbf{x}}\bullet\right)$ can be identified as the shift operator for the case of functions of many variables. This is the multivari-ate analogue of the translation (shift) operator $\exp(\hat{x}\, d\bullet/dx)$ for functions of one variable, first introduced by Lagrange; see also (Glaeske, Prudnikov and Skòrnik, 2006, Sec. 1.1).

Denoting the first component of $\mathbf{x}$ by (the scalar) $\upsilon$, and the remaining ones with the $N-$dimensional vector $\mathbf{u}$, Eq. (11) is recast as

$$G(\upsilon\,;\mathbf{u}) = G(\upsilon_0+\hat{\upsilon}\,;\mathbf{u}_0+\hat{\mathbf{u}}) = \mathcal{T}_{\hat{\upsilon}\hat{\mathbf{u}}}\, G(\upsilon_0\,;\mathbf{u}_0) = \\ = \exp\left(\hat{\upsilon}\,\frac{\partial}{\partial\upsilon} + \sum_{n=1}^{N}\hat{u}_n\,\frac{\partial}{\partial u_n}\right) G(\upsilon_0\,;\mathbf{u}_0) = \exp\left(\hat{\upsilon}\,\frac{\partial}{\partial\upsilon} + \hat{\mathbf{u}}\cdot\nabla_{\mathbf{u}}\right) G(\upsilon_0\,;\mathbf{u}_0), \tag{12}$$

providing us with the shift operator for a function of a scalar and a vector argument, $G(\upsilon;\mathbf{u})$. Now, by setting in the above equation $u_n = u(t_n)$, where $u(t)$ is a continuous function, and applying Volterra's approach to pass from the discrete to continuous (see Appendix A), we observe that the function $G(\upsilon\,;\mathbf{u})$ tends to the function-functional $\mathcal{G}[\upsilon\,;u(\bullet|_{t_0}^{t})]$ and the sum $\sum_{n=1}^{N}\hat{u}_n\,\partial/\partial u_n$ tends to the integral $\int_{t_0}^{t}\hat{u}(s)\,\delta/\delta u(s)\,ds$, where $\delta/\delta u(s)$ is the Volterra functional derivative. Thus, Eq. (12) suggests the following exponential form of the shift operator $\mathcal{T}_{\hat{\upsilon}\hat{u}(\cdot)}\bullet$ when applied to the function-functional $\mathcal{G}[\upsilon\,;u(\bullet|_{t_0}^{t})]$:

$$\mathcal{G}[\upsilon\,;u(\bullet|_{t_0}^{t})] = \mathcal{G}[\upsilon_0+\hat{\upsilon}\,;u_0(\bullet|_{t_0}^{t})+\hat{u}(\bullet|_{t_0}^{t})] = \\ = \mathcal{T}_{\hat{\upsilon}\hat{u}(\cdot)}\,\mathcal{G}[\upsilon_0\,;u_0(\bullet|_{t_0}^{t})] = \\ = \exp\left(\hat{\upsilon}\,\frac{\partial}{\partial\upsilon} + \int_{t_0}^{t}\hat{u}(s)\,\frac{\delta}{\delta u(s)}\,ds\right)\mathcal{G}[\upsilon_0\,;u_0(\bullet|_{t_0}^{t})]. \tag{13}$$

By omitting the scalar argument $\upsilon$, Eq. (13) defines the functional shift operator $\mathcal{T}_{\hat{u}(\cdot)}\bullet$, derived in (Sobczyk 1985, chap.I, Eq. 1.116'), where Volterra's passing was also employed. A rigorous derivation of Eq. (13) can be obtained by direct use of the Volterra-Taylor series expansion of $\mathcal{G}[\upsilon\,;u(\bullet|_{t_0}^{t})]$; see Appendix A.

## 4. Proof of Theorem 1

Equation (13) is the essential deterministic prerequisite for the proof of Theorem 1. Having it at our disposal, the proof of Theorem 1 is straightforward. Substituting, in Eq. (13), the arguments $\upsilon, u(\bullet)$ by the random arguments $X_0(\theta)$, $\Xi(\bullet;\theta)$, we obtain the following representation of the random RFF$\ell$ $\mathcal{G}[X_0(\theta)\,;\Xi(\bullet|_{t_0}^{t};\theta)]$:

$$\mathcal{G}[X_0(\theta)\,;\Xi(\bullet|_{t_0}^{t};\theta)] \equiv \mathcal{G}[m_{X_0}+\hat{X}_0(\theta)\,;m_{\Xi(\cdot)}(\bullet|_{t_0}^{t})+\hat{\Xi}(\bullet|_{t_0}^{t};\theta)] \equiv \\ \equiv \mathcal{T}_{\hat{X}_0\hat{\Xi}(\cdot)}(\theta)\left[\mathcal{G}[m_{X_0}\,;m_{\Xi(\cdot)}(\bullet|_{t_0}^{t})]\right] =$$

$$= \exp\left( \hat{X}_0(\theta)\,\frac{\partial}{\partial \upsilon} + \int_{t_0}^{t} ds\, \hat{\Xi}(s;\theta)\,\frac{\delta}{\delta u(s)} \right) \mathcal{G}[m_{X_0}; m_{\Xi(\cdot)}(\bullet|_{t_0}^{t})]. \tag{14}$$

Recall that $m_{X_0}$, $m_{\Xi(\cdot)}(\bullet)$ are the mean values, and $\hat{X}_0(\theta)$, $\hat{\Xi}(\bullet;\theta)$ are the fluctuations of the random elements $X_0(\theta)$, $\Xi(\bullet;\theta)$ around their mean values; see Eq. (4).

Eq. (14) provides us with a decomposition of the random RFF$\ell$ $\mathcal{G}[X_0(\theta);\Xi(\bullet;\theta)]$ in the form of a *random shift operator* "times" the deterministic function-functional $\mathcal{G}[m_{X_0}; m_{\Xi(\cdot)}(\bullet)]$. By averaging both sides of Eq. (14), we obtain

$$\mathbb{E}^{\theta}\left[\mathcal{G}[X_0(\theta);\Xi(\bullet|_{t_0}^{t};\theta)]\right] =$$

$$= \mathbb{E}^{\theta}\left[\exp\left( \hat{X}_0(\theta)\,\frac{\partial}{\partial \upsilon} + \int_{t_0}^{t} ds\, \hat{\Xi}(s;\theta)\,\frac{\delta}{\delta u(s)} \right)\right] \mathcal{G}[m_{X_0}; m_{\Xi(\cdot)}(\bullet|_{t_0}^{t})]. \tag{15}$$

The term $\mathbb{E}^{\theta}\left[\exp(\cdots)\right]$, in the right-hand side of Eq. (15), is called the *averaged shift operator*. Finally, recalling the form of the j.Ch.FF$\ell$, Eq. (2), we see that Eq. (15) can be also written as

$$\mathbb{E}^{\theta}\left[\mathcal{G}[X_0(\theta);\Xi(\bullet|_{t_0}^{t};\theta)]\right] = \varphi_{\hat{X}_0\hat{\Xi}(\cdot)}\left[\frac{\partial}{i\,\partial \upsilon};\frac{\delta}{i\,\delta u(s)}\right] \mathcal{G}[m_{X_0}; m_{\Xi(\cdot)}(\bullet|_{t_0}^{t})],$$

which is exactly Eq. (3). Thus, the proof of Theorem 1 is completed. In addition, the term $\varphi_{\hat{X}_0\hat{\Xi}(\cdot)}\left[\partial/i\,\partial\upsilon\,;\,\delta/i\,\delta u(s)\right]$ is identified as the averaged shift operator for RFF$\ell$s.

## 5. Proof of the extended NF Theorem

Since, in Theorem 2, the arguments $X_0(\theta)$, $\Xi(s;\theta)$ are assumed jointly Gaussian, their fluctuations $\hat{X}_0(\theta) = X_0(\theta) - m_{X_0}$, $\hat{\Xi}(s;\theta) = \Xi(s;\theta) - m_{\Xi(\cdot)}(s)$ are also jointly Gaussian with zero mean values and the same central moments $C_{\hat{X}_0\hat{X}_0} = C_{X_0X_0}$, $C_{\hat{X}_0\hat{\Xi}(\cdot)}(\bullet) = C_{X_0\Xi(\cdot)}(\bullet)$ and $C_{\hat{\Xi}(\cdot)\hat{\Xi}(\cdot)}(\bullet,\bullet) = C_{\Xi(\cdot)\Xi(\cdot)}(\bullet,\bullet)$. Thus, the j.Ch.FF$\ell$ of the random fluctuations around mean values, $\varphi_{\hat{X}_0\hat{\Xi}(\cdot)}[\upsilon; u(\bullet|_{t_0}^{t})]$, is now specified, by using Eq. (6), as

$$\varphi^{\text{Gauss}}_{\hat{X}_0\hat{\Xi}(\cdot)}[\upsilon; u(\bullet|_{t_0}^{t})] = \exp\left( -\frac{1}{2}\int_{t_0}^{t}\int_{t_0}^{t} C_{\Xi(\cdot)\Xi(\cdot)}(s_1,s_2)\,u(s_1)\,u(s_2)\,ds_1 ds_2 \right) \times$$

$$\times \exp\left(-\frac{1}{2}\,C_{X_0X_0}\,\upsilon^2\right)\cdot \exp\left(-\upsilon\int_{t_0}^{t} C_{X_0\Xi(\cdot)}(s)\,u(s)\,ds\right). \tag{16}$$

Accordingly, in this case, the averaged shift operator $\varphi_{\hat{X}_0\hat{\Xi}(\bullet)}\left[\partial/i\,\partial\upsilon\,;\,\delta/i\,\delta u(s)\right]$ can be expressed as the product of three operators

$$\varphi_{\hat{X}_0\hat{\Xi}(\bullet)}\left[\frac{\partial\bullet}{i\,\partial\upsilon}\,;\,\frac{\delta\bullet}{i\,\delta u(s)}\right] = \left(\bar{\mathcal{T}}_{\hat{X}_0\hat{X}_0}\bullet\right)\left(\bar{\mathcal{T}}_{\hat{X}_0\hat{\Xi}(\bullet)}\bullet\right)\left(\bar{\mathcal{T}}_{\hat{\Xi}(\bullet)\hat{\Xi}(\bullet)}\bullet\right), \tag{17}$$

defined by

$$\bar{\mathcal{T}}_{\hat{X}_0\hat{X}_0}\bullet = \exp\left(\frac{1}{2}\,C_{X_0X_0}\,\frac{\partial^2\bullet}{\partial\upsilon^2}\right), \tag{18a}$$

$$\bar{\mathcal{T}}_{\hat{X}_0\hat{\Xi}(\bullet)}\bullet = \exp\left(\int_{t_0}^{t} C_{X_0\Xi(\bullet)}(s)\,\frac{\partial\bullet}{\partial\upsilon}\,\frac{\delta\bullet}{\delta u(s)}\,ds\right), \tag{18b}$$

$$\bar{\mathcal{T}}_{\hat{\Xi}(\bullet)\hat{\Xi}(\bullet)}\bullet = \exp\left(\frac{1}{2}\int_{t_0}^{t}\int_{t_0}^{t} C_{\Xi(\bullet)\Xi(\bullet)}(s_1,\,s_2)\,\frac{\delta^2\bullet}{\delta u(s_1)\,\delta u(s_2)}\,ds_1ds_2\right). \tag{18c}$$

Operators $\bar{\mathcal{T}}$ can be considered as second-order versions of the shift operators, defined in Section 3. Thus, they will be called *quadratic averaged shift operators*. Before proceeding further (to complete the proof of Theorem 2), we shall study the basic properties of the said operators.

**Properties of operators $\bar{\mathcal{T}}$.** On $C^\infty$ function-functionals, operators $\bar{\mathcal{T}}$ are well-defined and they have the following properties, which are needed subsequently for the proof of the extended NF theorem.

**Lemma 1: Operators $\bar{\mathcal{T}}$ are linear.**

That is, for any two $C^\infty$ function-functionals, $\mathcal{G}[\upsilon\,;\,u(\bullet|_{t_0}^{t})]$, $\mathcal{F}[\upsilon\,;\,u(\bullet|_{t_0}^{t})]$ it holds true that

$$\bar{\mathcal{T}}\left[\alpha\,\mathcal{G}[\upsilon\,;\,u(\bullet|_{t_0}^{t})] + \beta\,\mathcal{F}[\upsilon\,;\,u(\bullet|_{t_0}^{t})]\right] = \alpha\,\bar{\mathcal{T}}\left[\mathcal{G}[\upsilon\,;\,u(\bullet|_{t_0}^{t})]\right] + \beta\,\bar{\mathcal{T}}\left[\mathcal{F}[\upsilon\,;\,u(\bullet|_{t_0}^{t})]\right], \tag{19}$$

where $\bar{\mathcal{T}}\bullet$ stands for any of the three operators $\bar{\mathcal{T}}_{\hat{X}_0\hat{X}_0}\bullet$, $\bar{\mathcal{T}}_{\hat{X}_0\hat{\Xi}(\bullet)}\bullet$, $\bar{\mathcal{T}}_{\hat{\Xi}(\bullet)\hat{\Xi}(\bullet)}\bullet$, and $\alpha$, $\beta$ are scalars or scalar functions having argument(s) different than the differentiation argument(s) appearing in the corresponding operator $\bar{\mathcal{T}}\bullet$.

**Lemma 2: Operators $\bar{\mathcal{T}}$ commute with $\upsilon-$ and $u(s)-$ differentiations.**

That is, for a $C^\infty$ function-functional $\mathcal{G}[\upsilon\,;\,u(\bullet|_{t_0}^{t})]$, and for $\bar{\mathcal{T}} \in \left\{\bar{\mathcal{T}}_{\hat{X}_0\hat{X}_0},\,\bar{\mathcal{T}}_{\hat{X}_0\hat{\Xi}(\bullet)},\,\bar{\mathcal{T}}_{\hat{\Xi}(\bullet)\hat{\Xi}(\bullet)}\right\}$,

$$\frac{\partial}{\partial\upsilon}\left[\bar{\mathcal{T}}\,\mathcal{G}[\upsilon\,;\,u(\bullet|_{t_0}^{t})]\right] = \bar{\mathcal{T}}\left[\frac{\partial\mathcal{G}[\upsilon\,;\,u(\bullet|_{t_0}^{t})]}{\partial\upsilon}\right], \tag{20a}$$

and

$$\frac{\delta}{\delta u(s)}\left[\bar{\mathcal{T}}\,\mathcal{G}[\upsilon\,;u(\bullet|_{t_0}^{t})]\right] = \bar{\mathcal{T}}\left[\frac{\delta\mathcal{G}[\upsilon\,;u(\bullet|_{t_0}^{t})]}{\delta u(s)}\right]. \tag{20b}$$

**Lemma 3: Operators $\bar{\mathcal{T}}$ commute with each other.**

That is, for any $C^{\infty}$ function-functional $\mathcal{G}[\upsilon\,;u(\bullet|_{t_0}^{t})]$

$$\bar{\mathcal{T}}_{\hat{X}_0\hat{X}_0}\,\bar{\mathcal{T}}_{\hat{X}_0\hat{\Xi}(\cdot)}\,\bar{\mathcal{T}}_{\hat{\Xi}(\cdot)\hat{\Xi}(\cdot)}\;\mathcal{G}[\upsilon\,;u(\bullet|_{t_0}^{t})] = \bar{\mathcal{T}}_{\hat{X}_0\hat{\Xi}(\cdot)}\,\bar{\mathcal{T}}_{\hat{X}_0\hat{X}_0}\,\bar{\mathcal{T}}_{\hat{\Xi}(\cdot)\hat{\Xi}(\cdot)}\;\mathcal{G}[\upsilon\,;u(\bullet|_{t_0}^{t})] = \cdots$$

$$\cdots = \bar{\mathcal{T}}_{\hat{\Xi}(\cdot)\hat{\Xi}(\cdot)}\,\bar{\mathcal{T}}_{\hat{X}_0\hat{X}_0}\,\bar{\mathcal{T}}_{\hat{X}_0\hat{\Xi}(\cdot)}\;\mathcal{G}[\upsilon\,;u(\bullet|_{t_0}^{t})]. \tag{21}$$

In other words, the product of the three operators $\bar{\mathcal{T}}$, under any permutation of their order, has the same action on $\mathcal{G}[\upsilon\,;u(\bullet|_{t_0}^{t})]$. Proofs of Lemmata 1-3 are presented in Appendix B.

Using Eqs. (17) and (18), Eq. (3), for jointly Gaussian arguments, takes the form

$$\mathbb{E}^{\theta}\left[\mathcal{G}[X_0(\theta)\,;\Xi(\bullet|_{t_0}^{t}\,;\theta)]\right] = \left[\bar{\mathcal{T}}_{\hat{\Xi}(\cdot)\hat{\Xi}(\cdot)}\,\bar{\mathcal{T}}_{\hat{X}_0\hat{\Xi}(\cdot)}\,\bar{\mathcal{T}}_{\hat{X}_0\hat{X}_0}\;\mathcal{G}[\upsilon\,;u(\bullet|_{t_0}^{t})]\right]_{\substack{\upsilon=m_{X_0}\\ u(\bullet)=m_{\Xi(\cdot)}(\bullet)}} \tag{22}$$

For the case $\mathcal{G}[\cdots] = \Xi(t\,;\theta)\,\mathcal{F}[\cdots]$, Eq. (5), as assumed in Theorem 2, Eq. (22) reads

$$\mathbb{E}^{\theta}\left[\Xi(t\,;\theta)\,\mathcal{F}[\cdots]\right] = \\ = \left\{\bar{\mathcal{T}}_{\hat{\Xi}(\cdot)\hat{\Xi}(\cdot)}\,\bar{\mathcal{T}}_{\hat{X}_0\hat{\Xi}(\cdot)}\,\bar{\mathcal{T}}_{\hat{X}_0\hat{X}_0}\left[u(t)\,\mathcal{F}[\upsilon\,;u(\bullet|_{t_0}^{t})]\right]\right\}_{\substack{\upsilon=m_{X_0}\\ u(\bullet)=m_{\Xi(\cdot)}(\bullet)}} \tag{23}$$

The main part of the proof for the extended NF theorem lies in the successive applications of the three operators $\bar{\mathcal{T}}$ on $u(t)\,\mathcal{F}[\upsilon\,;u(\bullet|_{t_0}^{t})]$. Since, by virtue of Lemma 3, the order of application of operators $\bar{\mathcal{T}}$ does not alter the final result, we choose the order shown in Eq. (23). The action of each operator $\bar{\mathcal{T}}$ is given by the following lemmata, the proofs of which are presented in Appendix B.

**Lemma 4.** The action of operator $\bar{\mathcal{T}}_{\hat{X}_0\hat{X}_0}$ on $u(t)\,\mathcal{F}[\upsilon\,;u(\bullet|_{t_0}^{t})]$ is given by

$$\bar{\mathcal{T}}_{\hat{X}_0\hat{X}_0}\left[u(t)\,\mathcal{F}[\upsilon\,;u(\bullet|_{t_0}^{t})]\right] = u(t)\,\bar{\mathcal{T}}_{\hat{X}_0\hat{X}_0}\left[\mathcal{F}[\upsilon\,;u(\bullet|_{t_0}^{t})]\right]. \tag{24}$$

In the right-hand side of Eq. (24), $\bar{\mathcal{T}}_{\hat{X}_0\hat{X}_0}\left[\mathcal{F}[\upsilon\,;u(\bullet)]\right]$ is a new function-functional, which is denoted by $\mathcal{F}_1[\upsilon\,;u(\bullet)]$. By applying operator $\bar{\mathcal{T}}_{\hat{X}_0\hat{\Xi}(\cdot)}$ on both sides of Eq. (24) we obtain

$$\bar{\mathcal{T}}_{\hat{X}_0\hat{\Xi}(\cdot)}\,\bar{\mathcal{T}}_{\hat{X}_0\hat{X}_0}\left[u(t)\,\mathcal{F}[\upsilon\,;u(\bullet|_{t_0}^{t})]\right] = \bar{\mathcal{T}}_{\hat{X}_0\hat{\Xi}(\cdot)}\left[u(t)\,\mathcal{F}_1[\upsilon\,;u(\bullet|_{t_0}^{t})]\right]. \tag{25}$$

The right-hand side of Eq. (25) is calculated by using the following result:

**Lemma 5.** The action of operator $\bar{\mathcal{T}}_{\hat{X}_0 \hat{\Xi}(\bullet)}$ on $u(t)\,\mathcal{F}_1[\upsilon\,;u(\bullet|_{t_0}^{t})]$ is given by

$$\bar{\mathcal{T}}_{\hat{X}_0 \hat{\Xi}(\bullet)}\left[u(t)\,\mathcal{F}_1[\upsilon\,;u(\bullet|_{t_0}^{t})]\right] = u(t)\,\bar{\mathcal{T}}_{\hat{X}_0 \hat{\Xi}(\bullet)}\left[\mathcal{F}_1[\upsilon\,;u(\bullet|_{t_0}^{t})]\right] + \\ + C_{X_0 \Xi(\bullet)}(t)\,\bar{\mathcal{T}}_{\hat{X}_0 \hat{\Xi}(\bullet)}\left[\frac{\partial \mathcal{F}_1[\upsilon\,;u(\bullet|_{t_0}^{t})]}{\partial \upsilon}\right]. \tag{26}$$

Denoting the term $\bar{\mathcal{T}}_{\hat{X}_0 \hat{\Xi}(\bullet)}\left[\mathcal{F}_1[\upsilon\,;u(\bullet)]\right]$ by $\mathcal{F}_2[\upsilon\,;u(\bullet)]$, and combining Eq. (26) and Eq. (25), we obtain

$$\bar{\mathcal{T}}_{\hat{X}_0 \hat{\Xi}(\bullet)}\,\bar{\mathcal{T}}_{\hat{X}_0 \hat{X}_0}\left[u(t)\,\mathcal{F}[\upsilon\,;u(\bullet|_{t_0}^{t})]\right] = u(t)\,\mathcal{F}_2[\upsilon\,;u(\bullet|_{t_0}^{t})] + \\ + C_{X_0 \Xi(\bullet)}(t)\,\bar{\mathcal{T}}_{\hat{X}_0 \hat{\Xi}(\bullet)}\left[\frac{\partial \mathcal{F}_1[\upsilon\,;u(\bullet|_{t_0}^{t})]}{\partial \upsilon}\right].$$

Applying the operator $\bar{\mathcal{T}}_{\hat{\Xi}(\bullet)\hat{\Xi}(\bullet)}$ on both sides of the above equation, leads to

$$\bar{\mathcal{T}}_{\hat{\Xi}(\bullet)\hat{\Xi}(\bullet)}\,\bar{\mathcal{T}}_{\hat{X}_0 \hat{\Xi}(\bullet)}\,\bar{\mathcal{T}}_{\hat{X}_0 \hat{X}_0}\left[u(t)\,\mathcal{F}[\upsilon\,;u(\bullet|_{t_0}^{t})]\right] = \bar{\mathcal{T}}_{\hat{\Xi}(\bullet)\hat{\Xi}(\bullet)}\left[u(t)\,\mathcal{F}_2[\upsilon\,;u(\bullet|_{t_0}^{t})]\right] + \\ + \bar{\mathcal{T}}_{\hat{\Xi}(\bullet)\hat{\Xi}(\bullet)}\left\{C_{X_0 \Xi(\bullet)}(t)\,\bar{\mathcal{T}}_{\hat{X}_0 \hat{\Xi}(\bullet)}\left[\frac{\partial \mathcal{F}_1[\upsilon\,;u(\bullet|_{t_0}^{t})]}{\partial \upsilon}\right]\right\}. \tag{27}$$

We shall now elaborate on the two terms appearing in the right-hand side of Eq. (27). By using the linearity of $\bar{\mathcal{T}}_{\hat{\Xi}(\bullet)\hat{\Xi}(\bullet)}$ (Lemma 1) and the commutation of $\bar{\mathcal{T}}_{\hat{X}_0 \hat{X}_0}$ with the $\upsilon-$derivative (Lemma 2) in conjunction with the definition of $\mathcal{F}_1[\upsilon\,;u(\bullet)]$, the second term in the right-hand side of Eq. (27) is expressed in terms of the function-functional $\mathcal{F}[\upsilon\,;u(\bullet)]$ as follows:

$$\bar{\mathcal{T}}_{\hat{\Xi}(\bullet)\hat{\Xi}(\bullet)}\left\{C_{X_0 \Xi(\bullet)}(t)\,\bar{\mathcal{T}}_{\hat{X}_0 \hat{\Xi}(\bullet)}\left[\frac{\partial \mathcal{F}_1[\upsilon\,;u(\bullet|_{t_0}^{t})]}{\partial \upsilon}\right]\right\} = \\ = C_{X_0 \Xi(\bullet)}(t)\,\bar{\mathcal{T}}_{\hat{\Xi}(\bullet)\hat{\Xi}(\bullet)}\,\bar{\mathcal{T}}_{\hat{X}_0 \hat{\Xi}(\bullet)}\,\bar{\mathcal{T}}_{\hat{X}_0 \hat{X}_0}\left[\frac{\partial \mathcal{F}[\upsilon\,;u(\bullet|_{t_0}^{t})]}{\partial \upsilon}\right]. \tag{28}$$

Concerning the first term in the right-hand side of Eq. (27), we need the following Lemma:

**Lemma 6.** The action of operator $\bar{\mathcal{T}}_{\hat{\Xi}(\bullet)\hat{\Xi}(\bullet)}$ on $u(t)\,\mathcal{F}_2[\upsilon\,;u(\bullet|_{t_0}^{t})]$ is given by

$$\bar{\mathcal{T}}_{\hat{\Xi}(\bullet)\hat{\Xi}(\bullet)}\left[u(t)\,\mathcal{F}_2[\upsilon\,;u(\bullet|_{t_0}^{t})]\right] = u(t)\,\bar{\mathcal{T}}_{\hat{\Xi}(\bullet)\hat{\Xi}(\bullet)}\left[\mathcal{F}_2[\upsilon\,;u(\bullet|_{t_0}^{t})]\right] + \\ + \int_{t_0}^{t} C_{\Xi(\bullet)\Xi(\bullet)}(t,s)\,\bar{\mathcal{T}}_{\hat{\Xi}(\bullet)\hat{\Xi}(\bullet)}\left[\frac{\delta\mathcal{F}_2[\upsilon\,;u(\bullet|_{t_0}^{t})]}{\delta u(s)}\right] ds. \qquad (29)$$

Having now at our disposal the results of the previous lemmata, we are able to proceed to the

*Finalization of the proof of Theorem* 2: Substituting Eqs. (28) and (29) into Eq. (27), recalling that $\mathcal{F}_2[\upsilon\,;u(\bullet)] = \bar{\mathcal{T}}_{\hat{X}_0\hat{\Xi}(\bullet)}\left[\mathcal{F}_1[\upsilon\,;u(\bullet)]\right]$ and $\mathcal{F}_1[\upsilon\,;u(\bullet)] = \bar{\mathcal{T}}_{\hat{X}_0\hat{X}_0}\left[\mathcal{F}[\upsilon\,;u(\bullet)]\right]$, and employing the commutation of operators $\bar{\mathcal{T}}_{\hat{X}_0\hat{\Xi}(\bullet)}$, $\bar{\mathcal{T}}_{\hat{X}_0\hat{X}_0}$ with the Volterra derivative $\delta\bullet/\delta u(s)$ (Lemma 2), we obtain

$$\bar{\mathcal{T}}_{\hat{\Xi}(\bullet)\hat{\Xi}(\bullet)}\bar{\mathcal{T}}_{\hat{X}_0\hat{\Xi}(\bullet)}\bar{\mathcal{T}}_{\hat{X}_0\hat{X}_0}\left[u(t)\,\mathcal{F}[\upsilon\,;u(\bullet|_{t_0}^{t})]\right] = \\ = u(t)\,\bar{\mathcal{T}}_{\hat{\Xi}(\bullet)\hat{\Xi}(\bullet)}\bar{\mathcal{T}}_{\hat{X}_0\hat{\Xi}(\bullet)}\bar{\mathcal{T}}_{\hat{X}_0\hat{X}_0}\left[\mathcal{F}[\upsilon\,;u(\bullet|_{t_0}^{t})]\right] + \\ + C_{X_0\Xi(\bullet)}(t)\,\bar{\mathcal{T}}_{\hat{\Xi}(\bullet)\hat{\Xi}(\bullet)}\bar{\mathcal{T}}_{\hat{X}_0\hat{\Xi}(\bullet)}\bar{\mathcal{T}}_{\hat{X}_0\hat{X}_0}\left[\frac{\partial\mathcal{F}[\upsilon\,;u(\bullet|_{t_0}^{t})]}{\partial\upsilon}\right] + \\ + \int_{t_0}^{t} C_{\Xi(\bullet)\Xi(\bullet)}(t,s)\,\bar{\mathcal{T}}_{\hat{\Xi}(\bullet)\hat{\Xi}(\bullet)}\bar{\mathcal{T}}_{\hat{X}_0\hat{\Xi}(\bullet)}\bar{\mathcal{T}}_{\hat{X}_0\hat{X}_0}\left[\frac{\delta\mathcal{F}[\upsilon\,;u(\bullet|_{t_0}^{t})]}{\delta u(s)}\right] ds. \qquad (30)$$

By setting $\upsilon = m_{X_0}$, and $u(\bullet) = m_{\Xi(\bullet)}(\bullet)$ in Eq. (30), and applying Eq. (22) to each term of the form $\bar{\mathcal{T}}_{\hat{\Xi}(\bullet)\hat{\Xi}(\bullet)}\bar{\mathcal{T}}_{\hat{X}_0\hat{\Xi}(\bullet)}\bar{\mathcal{T}}_{\hat{X}_0\hat{X}_0}[\cdots]$ in both sides of Eq. (30), we obtain the extended NF theorem, Eq. (7). The proof is now completed.

## 6. Application of the extended NF Theorem to the study of random differential equations under coloured Gaussian excitation

Let us consider the jointly Gaussian $X_0(\theta)$ and $\Xi(t;\theta)$ to be the initial value and the excitation to the nonlinear random differential equation (RDE):

$$\dot{X}(t;\theta) = h(X(t;\theta)) + \kappa\Xi(t;\theta), \qquad X(t_0;\theta) = X_0(\theta), \qquad (31a,b)$$

where $h(x)$ is a deterministic, continuous, nonlinear function, and $\kappa$ is a constant. The response pdf, $f_{X(t)}(x)$, of RDE (31) can be represented as the average of a random delta function (Hänggi, 1978; Sancho and San Miguel, 1980; Cetto and de la Peña, 1984; Fox, 1986; Venturi *et al.*, 2012; Wang, Tartakovsky and Tartakovsky, 2013; Athanassoulis, Kapelonis and Mamis, 2018):

$$f_{X(t)}(x) = \mathbb{E}^{\theta}\left[\delta(x - X(t;\theta))\right]. \qquad (32)$$

By differentiating both sides of Eq. (32) with respect to time and employing RDE (30), we obtain an evolution equation for $f_{X(t)}(x)$

$$\frac{\partial f_{X(t)}(x)}{\partial t} + \frac{\partial}{\partial x}\left(h(x)\, f_{X(t)}(x)\right) = -\kappa \frac{\partial}{\partial x}\left(\mathbb{E}^{\theta}\left[\Xi(t;\theta)\,\delta(x - X(t;\theta))\right]\right), \tag{33}$$

called the corresponding stochastic Liouville equation (SLE). By treating the solution $X(t;\theta)$ of RDE (31) as a function of the initial value $X_0(\theta)$ and a functional on the excitation $\Xi(\bullet;\theta)$ over the time interval $[t_0, t]$, the averaged term appearing in the right-hand side of SLE (33) is expressed as $\mathbb{E}^{\theta}\left[\Xi(t;\theta)\,\delta\left(x - X[X_0(\theta);\Xi(\bullet|_{t_0}^{t};\theta)]\right)\right]$, which is a form suitable for the application for the extended NF theorem, Eq. (7), with

$$\mathcal{F}[X_0(\theta);\Xi(\bullet|_{t_0}^{t};\theta)] = \delta\left(x - X[X_0(\theta);\Xi(\bullet|_{t_0}^{t};\theta)]\right). \tag{34}$$

Under the use of the extended NF Theorem, and the chain rule for the derivatives of delta function, SLE (33) is transformed into

$$\frac{\partial f_{X(t)}(x)}{\partial t} + \frac{\partial}{\partial x}\left[\left(h(x) + \kappa m_{\Xi(\cdot)}(t)\right) f_{X(t)}(x)\right] =$$

$$= \kappa\, C_{X_0\Xi(\cdot)}(t)\,\frac{\partial^2}{\partial x^2}\,\mathbb{E}^{\theta}\left[\delta(x - X(t;\theta))\,\frac{\partial X[X_0(\theta);\Xi(\bullet|_{t_0}^{t};\theta)]}{\partial X_0(\theta)}\right] +$$

$$+ \kappa \frac{\partial^2}{\partial x^2}\int_{t_0}^{t} C_{\Xi(\cdot)\Xi(\cdot)}(t, s)\,\mathbb{E}^{\theta}\left[\delta(x - X(t;\theta))\,\frac{\delta X[X_0(\theta);\Xi(\bullet|_{t_0}^{t};\theta)]}{\delta \Xi(s;\theta)}\right] ds. \tag{35}$$

For the case of zero mean excitation ($m_{\Xi(\cdot)}(t) = 0$), uncorrelated to the initial value ($C_{X_0\Xi(\cdot)}(t)$), transformed SLE (35) reduces to the SLE derived in (Sancho *et al.*, 1982; Hänggi *et al.*, 1985; Fox, 1987; Peacock-López, West and Lindenberg, 1988; Hänggi and Jung, 1995).

By employing a current-time approximation scheme (see references below) for the averaged terms in the right-hand side of the transformed SLE (35), we can obtain an approximate, yet closed equation governing the evolution of response pdf $f_{X(t)}(x)$. Such pdf evolution equations are commonly called generalized Fokker-Planck-Kolmogorov (genFPK) equations, since they constitute the counterpart of the classical Fokker-Planck-Kolmogorov equation for RDEs under coloured noise excitation. The transformed SLE (35) is preferred over the original SLE (33) as a starting point for the formulation of genFPK equations, since:

**i)** In transformed SLE (35) the mean effect of excitation is explicitly present in the drift coefficient (the additional term $\kappa m_{\Xi(\cdot)}(t)$ which is hidden in the original SLE (33)), **ii)** Inside the averaged terms of SLE (35), the derivatives of the response with respect to initial value $\partial X[X_0(\theta);\Xi(\bullet;\theta)]/\partial X_0(\theta)$ and excitation $\delta X[X_0(\theta);\Xi(\bullet;\theta)]/\delta\Xi(s;\theta)$ appear. These derivatives can be expressed via the solution of the variational problem associated with the nonlinear RDE (31). Since the variational problem is always linear, the said derivatives can be easily expressed in the following closed form for any RDE; see e.g. (Mamis, Athanassoulis and Papadopoulos, 2018),

$$\frac{\partial X[X_0(\theta);\Xi(\bullet|_{t_0}^{t};\theta)]}{\partial X_0(\theta)} = \exp\left(\int_{t_0}^{t} h'\big(X(u;\theta)\big)\,du\right), \tag{36a}$$

$$\frac{\delta X[X_0(\theta);\Xi(\bullet|_{t_0}^{t};\theta)]}{\delta\Xi(s;\theta)} = \kappa\exp\left(\int_{s}^{t} h'\big(X(u;\theta)\big)\,du\right), \tag{36b}$$

with the prime denoting the derivative of the function with respect to its argument. In the literature, various current-time approximations have been proposed for the transformed SLE, the most widely-used being the small correlation time (Sancho *et al.*, 1982), Hänggi's ansatz (Hänggi *et al.*, 1985) and Fox's approximation (Fox, 1986). In our recent works (Athanassoulis, Kapelonis and Mamis, 2018; Mamis, Athanassoulis and Kapelonis, 2018), a new approximation scheme, based on random Volterra-Taylor expansion series, was employed for the closure of Eq. (35). This approximation led to a family of novel genFPK equations which, for the non-restrictive representation $h(x) = \sum_{k=1}^{N}\eta_k\, g_k(x)$ in Eq. (31), where $\eta_k$'s are constants, $g_1(x) = x$, modeling the linear part, and $g_k(x)$, $k = 2,\dots,N$ are nonlinear functions, reads

$$\frac{\partial f_{X(t)}(x)}{\partial t} + \frac{\partial}{\partial x}\Big[\big(h(x) + \kappa m_{\Xi(\cdot)}(t)\big)\, f_{X(t)}(x)\Big] =$$
$$= \frac{\partial^2}{\partial x^2}\left[\left(D_0^{\text{eff}}\Big[R_{h'(\cdot)}(\bullet|_{t_0}^{t}),t\Big] + \sum_{m=1}^{M} D_m^{\text{eff}}\Big[R_{h'(\cdot)}(\bullet|_{t_0}^{t}),t\Big]\sum_{|\boldsymbol{\alpha}|=m}\frac{\boldsymbol{\varphi}^{\boldsymbol{\alpha}}\big(x;\{R_{g_k'(\cdot)}(t)\}\big)}{\boldsymbol{\alpha}!}\right) f_{X(t)}(x)\right], \tag{37}$$

where $M\in\mathbb{N}$, denoting the order of the approximation used, $R_{h'(\cdot)}(s) = \mathbb{E}^{\theta}\big[h'\big(X(s;\theta)\big)\big]$, $R_{g_k'(\cdot)}(s) = \mathbb{E}^{\theta}\big[g_k'\big(X(s;\theta)\big)\big]$ are moments of the response, $\boldsymbol{\alpha}=(\alpha_2,\dots,\alpha_N)$ is a multi-index, and

$$\boldsymbol{\varphi}\big(x;\{R_{g_k'(\cdot)}(t)\}\big) = \Big(\varphi_2\big(x;R_{g_2'(\cdot)}(t)\big),\dots,\varphi_N\big(x;R_{g_N'(\cdot)}(t)\big)\Big), \tag{38a}$$

with

$$\varphi_k(x) = \eta_k\big(g_k'(x) - R_{g_k'(\cdot)}(t)\big). \tag{38b}$$

Last, coefficients $D_m^{\text{eff}}[R_{h'(\cdot)}(\bullet|_{t_0}^{t}),t]$, called the (ordered) generalized effective noise intensities, are

$$D_m^{\text{eff}}[R_{h'(\cdot)}(\bullet|_{t_0}^{t}),t] = \kappa \exp\left(\int_{t_0}^{t} R_{h'(\cdot)}(u)\,du\right) C_{X_0\Xi(\cdot)}(t)\,(t-t_0)^m + \\ + \kappa^2 \int_{t_0}^{t} \exp\left(\int_{s}^{t} R_{h'(\cdot)}(u)\,du\right) C_{\Xi(\cdot)\Xi(\cdot)}(t,s)\,(t-s)^m\,ds. \tag{39}$$

Eq. (37) has been solved numerically and validated via Monte Carlo simulations (Athanassoulis, Kapelonis and Mamis, 2018; Mamis, Athanassoulis and Kapelonis, 2018), for a wide range of values of the coloured noise correlation time (up to five times the relaxation time of the ODE (31)) and noise intensity. In all these cases, the results obtained by solving Eq. (37) are in excellent agreement with MC simulations both in the transient and in the long-time regime.

**Remark 3.** The formulation of genFPK equations for RDEs with coloured excitation correlated with the random initial value has not been derived before, to the best of our knowledge, and it is the topic of some recent works of ours (Athanassoulis, Kapelonis and Mamis, 2018; Mamis, Athanassoulis and Papadopoulos, 2018). Note that, it is a feature worth including for certain applications, as mentioned in (Hänggi, 1989, Sec. 9.4; Wio, Colet and San Miguel, 1989).

**Remark 4.** The observant reader has already pointed out the discrepancy between the requirement, in NF theorem, for the RFF$\ell$ being $C^\infty$ with respect to its random arguments, and the use of random delta function, Eq. (34), as the said RFF$\ell$. A rigorous treatment of this point requires the formulation of the results in the context of generalized random functions, see (Gelfand and Vilenkin, 1964) Chapter 3, and their proof by reduction to the dual space of $C^\infty$ spaces. This analysis involves Gelfand-Shilov spaces and will not be performed herein.

## 7. First example and validation: The exact response pdf evolution equation corresponding to a linear random differential equation

By considering $h(x) = \eta x$, Eq. (31) becomes the linear random differential equation:

$$\dot{X}(t;\theta) = \eta X(t;\theta) + \kappa \Xi(t;\theta), \qquad X(t_0;\theta) = X_0(\theta). \tag{40a,b}$$

Since $h(x)$ is linear, the variational derivatives, Eqs. (36a,b), are independent from the response; $\partial X[X_0(\theta);\Xi(\bullet|_{t_0}^{t};\theta)]/\partial X_0(\theta) = e^{\eta(t-t_0)}$, $\delta X[X_0(\theta);\Xi(\bullet|_{t_0}^{t};\theta)]/\delta \Xi(s;\theta) = \kappa e^{\eta(t-s)}$, and thus the transformed SLE (35) is, in fact, the *exact* response pdf evolution equation:

$$\frac{\partial f_{X(t)}(x)}{\partial t} + \frac{\partial}{\partial x}\Big[\big(\eta x + \kappa m_{\Xi(\cdot)}(t)\big) f_{X(t)}(x)\Big] = D^{\text{eff}}(t)\,\frac{\partial^2 f_{X(t)}(x)}{\partial x^2}, \tag{41}$$

where the term $D^{\text{eff}}(t)$, called the effective noise intensity, is given by

$$D^{\text{eff}}(t) = \kappa\, e^{\eta(t-t_0)}\, C_{X_0\Xi(\cdot)}(t) + \kappa^2 \int_{t_0}^{t} e^{\eta(t-s)}\, C_{\Xi(\cdot)\Xi(\cdot)}(t,s)\,ds, \tag{42}$$

along with the initial Gaussian condition:

$$f_{X(t_0)}(x) = f_{X_0}(x) = \frac{1}{\sqrt{2\pi\sigma_{X_0}^2}} \exp\left[ -\frac{1}{2} \frac{(x - m_{X_0})^2}{\sigma_{X_0}^2} \right]. \tag{43}$$

where $m_{X_0}$, $\sigma_{X_0}^2$ are the initial mean value and initial variance respectively. It is a well-known result that the response process of any linear system, with Gaussian initial distribution, to an additive Gaussian excitation (either coloured or white) is also a Gaussian process. Furthermore, the mean value $m_{X(\cdot)}(t)$ and variance $\sigma_{X(\cdot)}^2(t)$ of the response $X(t;\theta)$ can be determined as the solutions to the respective moment equations, derived directly from RDE (40) (see e.g. (Sun, 2006) ch. 8, or (Athanassoulis, Tsantili and Kapelonis, 2015)):

$$\dot{m}_{X(\cdot)}(t) = \eta\, m_{X(\cdot)}(t) + \kappa\, m_{\Xi(\cdot)}(t), \qquad m_{X(\cdot)}(t_0) = m_{X_0}, \tag{44a,b}$$

$$\dot{\sigma}_{X(\cdot)}^2(t) = 2\eta\sigma_{X(\cdot)}^2(t) + 2\kappa C_{X_0\Xi(\cdot)}(t) e^{\eta(t-t_0)} + 2\kappa^2 \int_{t_0}^{t} C_{\Xi(\cdot)\Xi(\cdot)}\big(t,u\big)\, e^{\eta(t-u)}\, du, \tag{45a}$$

$$\sigma_{X(\cdot)}^2(t_0) = \sigma_{X_0}^2, \tag{45b}$$

with dot denoting the time differentiation. Note that Eq. (45a) can expressed equivalently, using Eq. (42), as

$$\dot{\sigma}_{X(\cdot)}^2(t) = 2\eta\sigma_{X(\cdot)}^2(t) + 2D^{\text{eff}}(t). \tag{45a′}$$

Solving initial value problems (44), (45), results into

$$m_{X(\cdot)}(t) = m_{X_0} e^{\eta(t-t_0)} + \kappa \int_{t_0}^{t} m_{\Xi(\cdot)}(\tau)\, e^{\eta(t-\tau)}\, d\tau, \tag{46}$$

and

$$\sigma_{X(\cdot)}^2(t) = \sigma_{X_0}^2 e^{2\eta(t-t_0)} + 2\int_{t_0}^{t} D^{\text{eff}}(\tau)\, e^{2\eta(t-\tau)}\, d\tau. \tag{47}$$

Returning to Eq. (41), and following (Sun, 2006) sec. 6.3, this pdf evolution equation can be solved by employing the Fourier transform; $\hat{f}(y,t) = \int_{\mathbb{R}} e^{iyx} f_{X(t)}(x)\, dx$, which leads to

$$\frac{\partial \hat{f}(y,t)}{\partial t} = \eta y \frac{\partial \hat{f}(y,t)}{\partial y} + \Big( i\kappa m_{\Xi(\cdot)}(t)\, y - D^{\text{eff}}(t)\, y^2 \Big) \hat{f}(y,t), \tag{48a}$$

supplemented with the transformed initial condition

$$\hat{f}(y,t_0) = \exp\left( i m_{X_0}\, y - \frac{1}{2}\sigma_{X_0}^2\, y^2 \right). \tag{48b}$$

Initial value problem (48) is solved (Polyanin, Zaitsev and Moussiaux, 2001) sec. 4.1, by first determining the characteristic curve $w(y,t) = ye^{\eta t}$ as the solution of the characteristic equa-tion $dt = -dy/(\eta y)$. Then, we seek a solution of the form $g(w)\,\exp\left(\int_{t_0}^{t} h(w,t)\,dt\right)$, where $g(w)$ is a function of the characteristic curve, to be defined by the initial condition (48b), and $h(w,t) = i\kappa m_{\Xi(\bullet)}(t)\, we^{-\eta t} - D^{\text{eff}}(t)\, w^2 e^{-2\eta t}$, that is the coefficient multiplying $\hat{f}(y,t)$ in Eq. (48a), rewritten in terms of $w$, $t$. Finally, by returning to the original variables $y$, $t$, we obtain the solution

$$\hat{f}(y,t) = \exp\left[ i \left( m_{X_0} e^{\eta(t-t_0)} + \kappa \int_{t_0}^{t} m_{\Xi(\bullet)}(\tau)\, e^{\eta(t-\tau)}\, d\tau \right) y \right] \times$$
$$\times \exp\left[ -\frac{1}{2} \left( \sigma_{X_0}^2 e^{2\eta(t-t_0)} + 2\int_{t_0}^{t} D^{\text{eff}}(\tau)\, e^{2\eta(t-\tau)}\, d\tau \right) y^2 \right] \tag{49}$$

which, by employing the inverse Fourier transform, and utilizing the formulae (46), (47) for $m_{X(\bullet)}(t)$ and $\sigma^2_{X(\bullet)}(t)$, results in

$$f_{X(t)}(x) = \frac{1}{\sqrt{2\pi\sigma^2_{X(\bullet)}(t)}} \exp\left[ -\frac{1}{2} \frac{\left(x - m_{X(\bullet)}(t)\right)^2}{\sigma^2_{X(\bullet)}(t)} \right], \tag{50}$$

which is the expected Gaussian distribution.

**Remark 5.** The uniqueness of Gaussian solution (50) is ensured by the injectivity of Fourier transform for absolutely integrable functions, and the uniqueness of solution for transformed problem (48a,b), see (Polyanin, Zaitsev and Moussiaux, 2001), Sec. 10.1.2.

**Remark 6.** To sum up, in this section, we have proven that response pdf evolution Eq. (41), corresponding to the linear RDE (40a,b), reproduces the expected Gaussian solution, Eq. (50), correctly. This result also constitutes a first validation for the novel, extended NF theorem, Eq. (7), since the derivation of Eq. (41) is based solely on random delta formalism (32) and the extended NF theorem.

## 8. Generalization of Theorems 1 and 2 to the multidimensional case

In this section, we concisely present the generalizations of Theorems 1 and 2 for the case where $\mathbf{X}_0(\theta) \equiv \mathbf{X}^0(\theta)$ is an $N$–dimensional random vector and $\mathbf{\Xi}(\bullet;\theta)$ is a $K$–dimensional random function. Note that, in general, dimensions $N$ and $K$ do not necessarily agree. Since the proofs of the generalized theorems are similar to those of Theorems 1, 2, their details are omitted.

**Theorem 3:** Under appropriate smoothness assumptions, the mean value of the RFF$\ell$ $\mathcal{G}[\mathbf{X}^0(\theta)\,;\mathbf{\Xi}(\bullet;\theta)]$ is expressed as

$$\mathbb{E}^{\theta}\left[\mathcal{G}[\mathbf{X}^0(\theta)\,;\mathbf{\Xi}(\bullet;\theta)]\right] = \varphi_{\hat{\mathbf{X}}^0\hat{\mathbf{\Xi}}(\bullet)}\left( \frac{\partial}{i\,\partial\boldsymbol{\upsilon}}\,;\frac{\delta}{i\,\delta\mathbf{u}(\bullet)} \right) \mathcal{G}[\boldsymbol{\upsilon};\mathbf{u}(\bullet|_{t_0}^{t})]\Big|_{\substack{\boldsymbol{\upsilon}=\mathbf{m}_{\mathbf{X}^0} \\ \mathbf{u}(\bullet)=\mathbf{m}_{\mathbf{\Xi}(\bullet)}(\bullet)}} =$$

$$= \mathbb{E}^{\theta}\left[\exp\left(\sum_{n=1}^{N}\hat{X}_n^0(\theta)\,\frac{\partial}{\partial \upsilon_n} + \sum_{k=1}^{K}\int_{t_0}^{t} ds\,\hat{\Xi}_k(s;\theta)\,\frac{\delta}{\delta u_k(s)}\right)\right]\mathcal{G}[\boldsymbol{\upsilon};\mathbf{u}(\bullet|_{t_0}^{t})]\Big|_{\substack{\boldsymbol{\upsilon}=\mathbf{m}_{\mathbf{X}^0}\\ \mathbf{u}(\bullet)=\mathbf{m}_{\Xi(\bullet)}(\bullet)}} \quad (51)$$

where $\hat{\mathbf{X}}^0(\theta) = \mathbf{X}^0(\theta) - \mathbf{m}_{\mathbf{X}^0}$, $\hat{\boldsymbol{\Xi}}(s;\theta) = \boldsymbol{\Xi}(s;\theta) - \mathbf{m}_{\Xi(\bullet)}(s)$ are the fluctuations of the random elements $\mathbf{X}^0(\theta)$ and $\boldsymbol{\Xi}(s;\theta)$ around their mean values, and $\varphi_{\hat{\mathbf{X}}^0\hat{\Xi}(\bullet)}[\boldsymbol{\upsilon};\mathbf{u}(\bullet)]$ is the j.Ch.FF$\ell$ of the said fluctuations. ■

As in Theorem 1, Theorem 3 is proved by identifying the similarity between the average of the random shift operator $\mathcal{T}_{\hat{\mathbf{X}}^0\hat{\Xi}(\bullet)}(\theta)[\bullet]$ for the multidimensional case, defined by

$$\mathcal{G}[\mathbf{X}^0(\theta);\boldsymbol{\Xi}(\bullet|_{t_0}^{t};\theta)] \equiv \mathcal{G}[\mathbf{m}_{\mathbf{X}^0} + \hat{\mathbf{X}}^0(\theta);\mathbf{m}_{\Xi(\bullet)}(\bullet|_{t_0}^{t}) + \hat{\boldsymbol{\Xi}}(\bullet|_{t_0}^{t};\theta)] \equiv$$

$$\equiv \mathcal{T}_{\hat{\mathbf{X}}^0\hat{\Xi}(\bullet)}(\theta)\left[\mathcal{G}[\mathbf{m}_{\mathbf{X}^0};\mathbf{m}_{\Xi(\bullet)}(\bullet|_{t_0}^{t})]\right] =$$

$$= \exp\left(\sum_{n=1}^{N}\hat{X}_n^0(\theta)\,\frac{\partial}{\partial \upsilon_n} + \sum_{k=1}^{K}\int_{t_0}^{t} ds\,\hat{\Xi}_k(s;\theta)\,\frac{\delta}{\delta u_k(s)}\right)\mathcal{G}[\mathbf{m}_{\mathbf{X}^0};\mathbf{m}_{\Xi(\bullet)}(\bullet|_{t_0}^{t})], \quad (52)$$

and the definition relation of j.Ch.FF$\ell$

$$\varphi_{\hat{\mathbf{X}}^0\hat{\Xi}(\bullet)}[\boldsymbol{\upsilon};\mathbf{u}(\bullet)] = \mathbb{E}^{\theta}\left[\exp\left(i\sum_{n=1}^{N}\hat{X}_n^0(\theta)\,\upsilon_n + i\sum_{k=1}^{K}\int_{t_0}^{t}\hat{\Xi}_k(s;\theta)\,u_k(s)\,ds\right)\right]. \quad (53)$$

Eqs. (52), (53) can be obtained using Volterra's passing from the discrete to continuous, as has been done before for their scalar counterparts, Eqs. (14), (2) respectively.

**Theorem 4: the extended Novikov-Furutsu (NF) theorem for the multidimensional case.** Under appropriate smoothness assumptions, for the RFF$\ell$ $\mathcal{F}[\mathbf{X}^0(\theta);\boldsymbol{\Xi}(\bullet|_{t_0}^{t};\theta)] \equiv \mathcal{F}[\cdots]$, with jointly Gaussian arguments $\mathbf{X}^0(\theta)$, $\boldsymbol{\Xi}(\bullet;\theta)$, the following formulae hold true

$$\mathbb{E}^{\theta}\left[\Xi_k(t;\theta)\,\mathcal{F}[\cdots]\right] =$$

$$= m_{\Xi_k(\bullet)}(t)\,\mathbb{E}^{\theta}[\mathcal{F}[\cdots]] + \sum_{n=1}^{N} C_{X_n^0\Xi_k(\bullet)}(t)\,\mathbb{E}^{\theta}\left[\frac{\partial\mathcal{F}[\cdots]}{\partial X_n^0(\theta)}\right] + \quad (54)$$

$$+ \sum_{\ell=1}^{K}\int_{t_0}^{t} C_{\Xi_k(\bullet)\Xi_\ell(\bullet)}(t,s)\,\mathbb{E}^{\theta}\left[\frac{\delta\mathcal{F}[\cdots]}{\delta\Xi_\ell(s;\theta)}\right]ds. \quad ■$$

Eq. (54) is proved by substituting $\mathcal{G}[\mathbf{X}^0(\theta);\boldsymbol{\Xi}(\bullet|_{t_0}^{t};\theta)] = \Xi_k(t;\theta)\,\mathcal{F}[\cdots]$ in Eq. (51), and employing the j.Ch.FF$\ell$ of jointly Gaussian $\hat{\mathbf{X}}^0(\theta)$, $\hat{\boldsymbol{\Xi}}(\bullet;\theta)$:

$$\varphi^{\text{Gauss}}_{\hat{\mathbf{X}}^0\hat{\Xi}(\bullet)}[\boldsymbol{\upsilon}\,;\mathbf{u}(\bullet|_{t_0}^{t})] =$$

$$= \exp\left(-\frac{1}{2}\sum_{k_1=1}^{K}\sum_{k_2=1}^{K}\int_{t_0}^{t}\int_{t_0}^{t} C_{\Xi_{k_1}(\bullet)\Xi_{k_2}(\bullet)}(s_1, s_2)\, u_{k_1}(s_1)\, u_{k_2}(s_2)\, ds_1 ds_2\right)\times$$

$$\times \exp\left(-\frac{1}{2}\sum_{n_1=1}^{N}\sum_{n_2=1}^{N} C_{X^0_{n_1}X^0_{n_2}}\, \upsilon_{n_1}\upsilon_{n_2}\right)\cdot \exp\left(-\sum_{n=1}^{N}\upsilon_n\sum_{k=1}^{K}\int_{t_0}^{t} C_{X^0_n\Xi_k(\bullet)}(s)\, u_k(s)\, ds\right). \tag{55}$$

Eq. (55) can be obtained by using Volterra's approach. For this purpose, a procedure similar to the proof of Theorem 2 in Section 5 is followed, by defining and studying the appropriate operators $\bar{\mathcal{T}}$.

## 9. Summary and conclusions

In this work, an extension of the classical NF theorem for the case of RFF$\ell$s with jointly Gaussian arguments has been presented (Theorem 2). In fact, the said extension of the NF theorem is based on a representation of the mean value of any RFF$\ell$ via the j.Ch.FF$\ell$ of its (possibly non-Gaussian) arguments (Theorem 1). The latter, more general, result is an alternative to the series expansions for averaged random functionals given by (Bochkov, Dubkov and Malakhov, 1977). Extensions of the two aforementioned theorems to the multidimensional case have also been formulated as Theorem 4 and Theorem 3, respectively. In all cases, the random arguments assumed to have arbitrary mean values, and they can be mutually dependent.

Apart from the extensions, the proof of NF theorem presented herein is significantly different from the ones presented by other authors, and it is easily generalizable. The original proof by Novikov (Novikov, 1965), found also in (Ishimaru, 1978 Appendix 20B; Konotop and Vazquez, 1994; Scott, 2013) lies in expanding the functional $\mathcal{F}[\Xi(\bullet;\theta)]$ into a Volterra-Taylor series with respect to $\Xi(\bullet;\theta)$ around zero, multiplying both sides by $\Xi(t;\theta)$ and then taking the average. For the completion of the proof, Isserlis' theorem (Isserlis, 1918) for computing higher order moments in terms of the autocorrelation for the Gaussian case is invoked. Thus, Novikov's proof is essentially restricted to the Gaussian case. Our derivation is based on a generic result (Theorem 1), which is valid for any type of random arguments, and the specification of the characteristic (function-)functional as Gaussian, allowing for generalizations to other cases. In (Sobczyk, 1985; Klyatskin, 2005) the starting point of the proof is the functional shift operator in exponential form, as in the present work. However, a different operator formalism is used afterwards, making the proof less transparent, to the authors' opinion. Recall that all existing proofs refers to "pure" random functionals, while our results cover the case of random function-functionals, permitting the study of RDEs with random initial values, possibly correlated with coloured noise excitation(s).

## Appendix A. Volterra's principle of passing from discrete to continuous. Volterra calculus

A naïve, yet useful, way to describe Volterra's principle of passing from the discrete to continuous is the following:

Consider the discrete quantities $\Xi_n$, $u_n$, $n = 1(1)N$, obtained by sampling the continuous functions $\Xi(t)$, $u(t)$ over the interval $[a, b]$:

$$\Xi_n = \Xi(t_n), \qquad u_n = u(t_n), \qquad t_n \in [a, b].$$

Assume further that the points $t_n$ are ordered, $a \le t_1 < t_2 < \cdots < t_n < \cdots < t_N \le b$, and, as $N \to \infty$, all increments $\Delta t_n = t_{n+1} - t_n$, $n = 1(1)(N-1)$, $\Delta t_0 = t_1 - a$ and $\Delta t_N = b - t_N$, tend to zero. Under this assumption, Volterra's principle of passing from the discrete to continuous consists of the replacement of the sums

$$\sum_{n=1}^{N} \Xi_n = \sum_{n=1}^{N} \Xi(t_n), \qquad \sum_{n=1}^{N} \Xi_n \, u_n = \sum_{n=1}^{N} \Xi(t_n)\, u(t_n)$$

by

$$\sum_{n=1}^{N} \Xi(s_n)\, \Delta t_n, \qquad \sum_{n=1}^{N} \Xi(s_n)\, u(s_n)\, \Delta t_n, \text{ (2)}$$

and their interpretation (in the limit, as $N \to \infty$) as Riemann integrals,

$$\int_a^b \Xi(s)\, ds, \qquad \int_a^b \Xi(s)\, u(s)\, ds.$$

This line of thought is especially useful for the probabilistic study of random functions (in continuous time), since the primitive probabilistic information associated with a random function $\Xi(t;\theta)$ is the joint distribution (or the joint characteristic function) of the random variables $\Xi_n(\theta) = \Xi(t_n;\theta)$, where the time instances $t_1 < t_2 < \cdots < t_n < \cdots < t_N$ are distributed over the (common) domain of definition of path functions. For example, considering the joint characteristic function of the random variables $X_0(\theta)$, $\Xi(s_1;\theta)$, $\Xi(s_2;\theta)$, $\cdots$, $\Xi(s_N;\theta)$,

$$\varphi_{X_0 \Xi(s_1)\cdots\Xi(s_N)}\big(\upsilon; u(s_1), \ldots, u(s_N)\big) = \mathbb{E}^{\theta}\left[\exp\left(i X_0(\theta)\,\upsilon + i \sum_{n=1}^{N} \Xi(s_n;\theta)\, u(s_n)\, \Delta s\right)\right], \tag{A1}$$

and applying the above described Volterra's passing from the discrete to continuous, we obtain the j.Ch.FF$\ell$ of $X_0(\theta)$, $\Xi(t;\theta)$:

$$\varphi_{X_0 \Xi(\cdot)}[\upsilon; u(\bullet|_{t_0}^{t})] = \mathbb{E}^{\theta}_{\mathbf{P}_{X_0 \Xi(\cdot)}}\left[\exp\left(i X_0(\theta)\,\upsilon + i \int_{t_0}^{t} \Xi(s;\theta)\, u(s)\, ds\right)\right]. \tag{A2}$$

---

(2) $s_n \in [t_n, t_{n+1}]$. Any choice of $s_n$ leads to the same results, since the functions $\Xi(t)$, $u(t)$ are assumed to be continuous.

Another interesting application of Volterra's passing from the discrete to continuous is the construction of the j.Ch.FF$\ell$ of a jointly Gaussian scalar random variable and scalar random function, which is of fundamental importance for the proof of the extended NF theorem. In order to find this j.Ch.FF$\ell$ we begin from its discrete analogue, the joint characteristic function of a random variable $X_0(\theta)$ and a random vector $\boldsymbol{\Xi}(\theta)$. This is obtained by simple manipulations of a $(1+N)-$dimensional Gaussian characteristic function (see e.g. (Lukacs and Laha, 1964), Sec. 2.1), and reads as follows:

$$\varphi^{\text{Gauss}}_{X_0\boldsymbol{\Xi}}(\upsilon;\mathbf{u}) = \exp\left( i\sum_{n=1}^{N}\left(m_{\boldsymbol{\Xi}}\right)_n u_n - \frac{1}{2}\sum_{n=1}^{N}\sum_{m=1}^{N}\left(C_{\boldsymbol{\Xi}\boldsymbol{\Xi}}\right)_{nm} u_n u_m \right)\times$$
$$\times\exp\left( i m_{X_0}\,\upsilon - \frac{1}{2}C_{X_0X_0}\,\upsilon^2\right)\cdot\exp\left( -\upsilon\sum_{n=1}^{N}\left(C_{X_0\boldsymbol{\Xi}}\right)_n u_n\right), \tag{A3}$$

where $m_{X_0}$, $C_{X_0X_0}$ are the mean value and the variance of the random variable $X_0(\theta)$, $\mathbf{m}_{\boldsymbol{\Xi}}$, $\mathbf{C}_{\boldsymbol{\Xi}\boldsymbol{\Xi}}$ are the mean value and autocovariance of the random vector $\boldsymbol{\Xi}(\theta)$, and $\mathbf{C}_{X_0\boldsymbol{\Xi}}$ denotes their cross-covariance. By setting $\Xi_n(\theta)=\Xi(t_n;\theta)$ and $u_n=u(t_n)$ and considering the Volterra's limiting process as described above, we obtain the Gaussian j.Ch.FF$\ell$

$$\varphi^{\text{Gauss}}_{X_0\,\Xi(\bullet)}[\upsilon\,;u(\bullet|^{t}_{t_0})] =$$
$$= \exp\left( i\int_{t_0}^{t} m_{\Xi(\bullet)}(s)u(s)\,ds - \frac{1}{2}\int_{t_0}^{t}\int_{t_0}^{t} C_{\Xi(\bullet)\Xi(\bullet)}(s_1,s_2)\,u(s_1)\,u(s_2)\,ds_1 ds_2\right)\times$$
$$\times\exp\left( i m_{X_0}\,\upsilon - \frac{1}{2}C_{X_0X_0}\,\upsilon^2\right)\cdot\exp\left( -\upsilon\int_{t_0}^{t} C_{X_0\Xi(\bullet)}(s)\,u(s)\,ds\right). \tag{A4}$$

In Eq. (A4), we observe that $\varphi^{\text{Gauss}}_{X_0\,\Xi(\bullet)}[\upsilon\,;u(\bullet|^{t}_{t_0})]$ is the product of the Gaussian characteristic functional of $\Xi(\bullet;\theta)$, the Gaussian characteristic function of $X_0(\theta)$, as well as a term encapsulating their probabilistic dependence via their cross-correlation function $C_{X_0\Xi(\bullet)}(s)$.

Passing from discrete to continuous is only a minor trick of Volterra's approach. The vigor of the latter is amply revealed only when consider the corresponding functional calculus. Volterra functional derivative $\delta/\delta u(s)$ is the continuous analogue of the (usual) partial derivative $\partial/\partial u_n$, and leads to a well-structured calculus on functionals (Volterra, 1930; Volterra and Pérès, 1936; Averbukh and Smolyanov, 1967). An example illustrating the rigorousness of Volterra calculus, as well as the relation between Volterra functional derivative and the more widely known Gâteaux derivative, is the following derivation of the of Taylor's theorem for functionals, called the Volterra-Taylor theorem:

Let us consider the functional $\mathcal{G}[u(\bullet|_{t_0}^{t})]$, assumed to be $M+1$ times Gâteaux differentiable, with the goal of expanding it around $u_0(\bullet)$. By assuming $u(\bullet)$, $u_0(\bullet)$ as fixed, we define the real-valued function of one scalar variable $g(\varepsilon) = \mathcal{G}\left[u_0(\bullet|_{t_0}^{t}) + \varepsilon\delta u(\bullet|_{t_0}^{t})\right]$, where $\delta u(\bullet) = u(\bullet) - u_0(\bullet)$. Obviously, $g(\varepsilon)$ is well-defined and continuous in $[0,1]$, has the properties

$$g(0) = \mathcal{G}[u_0(\bullet|_{t_0}^{t})], \qquad \text{and} \qquad g(1) = \mathcal{G}[u(\bullet|_{t_0}^{t})],$$

and it is $M+1$ times differentiable, in the usual sense, in the interval $[0,1]$. Its Taylor expansion, for $g(1)$ around $g(0)$, reads as

$$g(1) = \sum_{m=0}^{M} \frac{1}{m!} \frac{d^m g(\varepsilon)}{d\varepsilon^m}\bigg|_{\varepsilon=0} + \frac{1}{(M+1)!} \frac{d^{M+1} g(\theta)}{d\varepsilon^{M+1}}, \qquad 0 < \theta < 1. \tag{A5}$$

Rewriting Eq. (A5) in terms of the functional $\mathcal{G}[u(\bullet|_{t_0}^{t})]$ we obtain

$$\begin{aligned} \mathcal{G}[u(\bullet|_{t_0}^{t})] = {} & \sum_{m=0}^{M} \frac{1}{m!} \frac{\partial^m}{\partial\varepsilon^m} \mathcal{G}\left[u_0(\bullet|_{t_0}^{t}) + \varepsilon\delta u(\bullet|_{t_0}^{t})\right]\bigg|_{\varepsilon=0} + \\ & + \frac{1}{(M+1)!} \frac{\partial^{M+1}}{\partial\varepsilon^{M+1}} \mathcal{G}\left[u_0(\bullet|_{t_0}^{t}) + \varepsilon\delta u(\bullet|_{t_0}^{t})\right]\bigg|_{\varepsilon=\theta}, \quad 0 < \theta < 1. \end{aligned} \tag{A6}$$

The terms $\frac{\partial^m}{\partial\varepsilon^m} \mathcal{G}\left[u_0(\bullet|_{t_0}^{t}) + \varepsilon\delta u(\bullet|_{t_0}^{t})\right]\Big|_{\varepsilon=0}$ are identified as the Gâteaux functional derivatives (Averbukh and Smolyanov, 1968), of $\mathcal{G}$ calculated at $u_0(\bullet)$ in the direction of $\delta u(\bullet)$, denoted by $\delta^m \mathcal{G}\left[u_0(\bullet|_{t_0}^{t}) ; \delta u(\bullet|_{t_0}^{t})\right]$. Thus, Eq. (A6) can be written as

$$\begin{aligned} \mathcal{G}[u(\bullet|_{t_0}^{t})] = {} & \sum_{m=0}^{M} \frac{1}{m!} \delta^m \mathcal{G}\left[u_0(\bullet|_{t_0}^{t}) ; \delta u(\bullet|_{t_0}^{t})\right] + \\ & + \frac{1}{(M+1)!} \frac{\partial^{M+1}}{\partial\varepsilon^{M+1}} \mathcal{G}\left[u_0(\bullet|_{t_0}^{t}) + \varepsilon\delta u(\bullet|_{t_0}^{t})\right]\bigg|_{\varepsilon=\theta}, \quad 0 < \theta < 1, \end{aligned} \tag{A7}$$

Eq. (A7) is the Gâteaux-Taylor series expansion for functionals. Under some technical assumptions (see (Hall, 1978; Hamilton, 1980)), it can be proved that Gâteaux derivatives are expressed in terms of Volterra derivatives by formulae

$$\delta\mathcal{G}\left[u_0(\bullet|_{t_0}^{t}) ; \delta u(\bullet|_{t_0}^{t})\right] = \int_{t_0}^{t} \frac{\delta\mathcal{G}[u_0(\bullet|_{t_0}^{t})]}{\delta u(s)} \delta u(s)\, ds,$$

for the first derivative and, in general,

$$\delta^m \mathcal{G}\Big[ u_0(\bullet|_{t_0}^{t}) \,;\, \delta u(\bullet|_{t_0}^{t}) \Big] = \int_{t_0}^{t} \overset{(m)}{\cdots} \int_{t_0}^{t} \frac{\delta^m \mathcal{G}[u_0(\bullet|_{t_0}^{t})]}{\delta u(s_1)\cdots \delta u(s_m)} \, \delta u(s_1)\cdots \delta u(s_m)\, ds_1 \cdots ds_m$$

for the $m^{\text{th}}$ derivative. Using the above equation, the Gâteaux-Taylor expansion, Eq. (A7), is re-written as a Volterra-Taylor expansion in the form

$$\mathcal{G}[u(\bullet|_{t_0}^{t})] = \sum_{m=0}^{M} \frac{1}{m!} \int_{t_0}^{t} \overset{(m)}{\cdots} \int_{t_0}^{t} \frac{\delta^m \mathcal{G}[u_0(\bullet|_{t_0}^{t})]}{\delta u(s_1)\cdots \delta u(s_m)} \, \delta u(s_1)\cdots \delta u(s_m)\, ds_1 \cdots ds_m \; + \quad \text{(A8)}$$

$$+ \big(\text{Residual term}\big)\,.$$

By considering now that $\mathcal{G}[u(\bullet|_{t_0}^{t})]$ is infinitely Gâteaux-differentiable, and the functional series in the right-hand side of Eq. (A8) converges, we obtain the infinite Volterra-Taylor series expansion of the functional,

$$\mathcal{G}[u(\bullet|_{t_0}^{t})] = \sum_{m=0}^{\infty} \frac{1}{m!} \int_{t_0}^{t} \overset{(m)}{\cdots} \int_{t_0}^{t} \frac{\delta^m \mathcal{G}[u_0(\bullet|_{t_0}^{t})]}{\delta u(s_1)\cdots \delta u(s_m)} \, \delta u(s_1)\cdots \delta u(s_m)\, ds_1 \cdots ds_m \,. \quad \text{(A9)}$$

The structure of Eq. (A9) is the same with the Taylor expansion for a function $G(\mathbf{u})$ of an $N-$ dimensional vector $\boldsymbol{u}$ around $\mathbf{u}_0$:

$$G(\mathbf{u}) = \sum_{m=0}^{\infty} \frac{1}{m!} \sum_{n_1=1}^{N} \overset{(m)}{\cdots} \sum_{n_m=1}^{N} \frac{\partial^m G(\mathbf{u}_0)}{\partial u_{n_1} \cdots \partial u_{n_m}} \, \delta u_{n_1} \cdots \delta u_{n_m} \,, \quad \text{(A10)}$$

where $\delta u_n = u_n - u_{0n}$. It is easy to see that Eq. (A9) can be also obtained from Eq. (A10) by applying the process of passing from the discrete to continuous, as described in the beginning of this section. The above analysis gives a clear indication of the validity of Volterra's principle (under some technical assumptions) and, in addition, establishes the fact the Volterra functional derivative is the continuous analogue of the (discrete) partial derivative.

### Appendix B. Proofs of Lemmata 1-6

As the proofs in (Jia, Tang and Kempf, 2017), where integration by differentiation is also performed, Lemmata 1-6 are proven using operators $\bar{\mathcal{T}}$ in series form, which are obtained by expanding the exponential in the right-hand sides of Eqs. (18a,b,c) of the main paper:

$$\bar{\mathcal{T}}_{\hat{X}_0 \hat{X}_0} \bullet = \sum_{p=0}^{\infty} \frac{1}{p!} \frac{1}{2^p} C_{X_0 X_0}^{p} \frac{\partial^{2p} \bullet}{\partial \upsilon^{2p}} \,, \quad \text{(B1a)}$$

$$\bar{\mathcal{T}}_{\hat{X}_0 \hat{\Xi}(\cdot)} \bullet =$$

$$= \sum_{p=0}^{\infty} \frac{1}{p!} \int_{t_0}^{t} \overset{(p)}{\cdots} \int_{t_0}^{t} C_{X_0 \Xi(\cdot)}(s^{(1)}) \cdots C_{X_0 \Xi(\cdot)}(s^{(p)}) \, \frac{\partial^p \, \delta^p \bullet}{\partial \upsilon^p \, \delta u(s^{(1)}) \cdots \delta u(s^{(p)})} \, ds^{(1)} \cdots ds^{(p)} \,, \quad \text{(B1b)}$$

$$\bar{\mathcal{T}}_{\hat{\Xi}(\bullet)\hat{\Xi}(\bullet)}\bullet = \sum_{p=0}^{\infty}\frac{1}{p!}\frac{1}{2^p}\int_{t_0}^{t}\overset{(2p)}{\cdots}\int_{t_0}^{t}\Big[C_{\Xi(\bullet)\Xi(\bullet)}(s_1^{(1)},s_2^{(1)})\cdots C_{\Xi(\bullet)\Xi(\bullet)}(s_1^{(p)},s_2^{(p)})\times \\ \times\frac{\delta^{2p}\,\bullet}{\delta u(s_1^{(1)})\delta u(s_2^{(1)})\cdots\delta u(s_1^{(p)})\delta u(s_2^{(p)})}\,ds_1^{(1)}ds_2^{(1)}\cdots ds_1^{(p)}ds_2^{(p)}\Big]. \tag{B1c}$$

The proofs of Lemmata 1-3 are based on the series expansions given above, in conjunction with the linearity of integrals and derivatives.

**Proof of Lemma 1: Operators $\bar{\mathcal{T}}$ are linear.** The action of operator $\bar{\mathcal{T}}_{\hat{X}_0\hat{\Xi}(\bullet)}$ on $\alpha\,\mathcal{G}[\upsilon\,;u(\bullet|_{t_0}^{t})]+\beta\,\mathcal{F}[\upsilon\,;u(\bullet|_{t_0}^{t})]$ is expressed via Eq. (B1b) as

$$\bar{\mathcal{T}}_{\hat{X}_0\hat{\Xi}(\bullet)}\Big[\alpha\,\mathcal{G}[\upsilon\,;u(\bullet|_{t_0}^{t})]+\beta\,\mathcal{F}[\upsilon\,;u(\bullet|_{t_0}^{t})]\Big]=$$

$$=\sum_{p=0}^{\infty}\frac{1}{p!}\int_{t_0}^{t}\overset{(p)}{\cdots}\int_{t_0}^{t}C_{X_0\Xi(\bullet)}(s^{(1)})\cdots C_{X_0\Xi(\bullet)}(s^{(p)})\,\frac{\partial^p\delta^p\Big[\alpha\,\mathcal{G}[\upsilon\,;u(\bullet|_{t_0}^{t})]+\beta\,\mathcal{F}[\upsilon\,;u(\bullet|_{t_0}^{t})]\Big]}{\partial\upsilon^p\;\delta u(s^{(1)})\cdots\delta u(s^{(p)})}ds^{(1)}\cdots ds^{(p)}.$$

Employing the linearity of derivatives and integrals, in conjunction with the assumption that $\alpha$, $\beta$ are independent from the differentiation arguments $\upsilon$ and $u(\bullet)$ of operator $\bar{\mathcal{T}}_{\hat{X}_0\hat{\Xi}(\bullet)}$, each term of the right-hand side of the above equation is linearly decomposed, resulting in the linearity of the operator $\bar{\mathcal{T}}_{\hat{X}_0\hat{\Xi}(\bullet)}$. Thus, Lemma 1 is proved for operator $\bar{\mathcal{T}}_{\hat{X}_0\hat{\Xi}(\bullet)}$. The proofs for $\bar{\mathcal{T}}_{\hat{X}_0\hat{X}_0}$, $\bar{\mathcal{T}}_{\hat{\Xi}(\bullet)\hat{\Xi}(\bullet)}$ are similar. ■

**Proof of Lemma 2: Operators $\bar{\mathcal{T}}$ commute with $\upsilon-$ and $u(s)-$ differentiations.** Using Eq. (B1b), we may write

$$\frac{\partial}{\partial\upsilon}\Big[\bar{\mathcal{T}}_{\hat{X}_0\hat{\Xi}(\bullet)}\,\mathcal{G}[\upsilon\,;u(\bullet|_{t_0}^{t})]\Big]=$$

$$=\frac{\partial}{\partial\upsilon}\sum_{p=0}^{\infty}\frac{1}{p!}\int_{t_0}^{t}\overset{(p)}{\cdots}\int_{t_0}^{t}C_{X_0\Xi(\bullet)}(s^{(1)})\cdots C_{X_0\Xi(\bullet)}(s^{(p)})\,\frac{\partial^p\,\delta^p\,\mathcal{G}[\upsilon\,;u(\bullet|_{t_0}^{t})]}{\partial\upsilon^p\;\delta u(s^{(1)})\cdots\delta u(s^{(p)})}\,ds^{(1)}\cdots ds^{(p)}$$

using the linearity and continuity of the derivative

$$=\sum_{p=0}^{\infty}\frac{1}{p!}\int_{t_0}^{t}\overset{(p)}{\cdots}\int_{t_0}^{t}C_{X_0\Xi(\bullet)}(s^{(1)})\cdots C_{X_0\Xi(\bullet)}(s^{(p)})\,\frac{\partial^{p+1}\,\delta^p\,\mathcal{G}[\upsilon\,;u(\bullet|_{t_0}^{t})]}{\partial\upsilon^{p+1}\;\delta u(s^{(1)})\cdots\delta u(s^{(p)})}\,ds^{(1)}\cdots ds^{(p)}$$

which is identified, via Eq. (B1b), as the infinite series of

$$=\bar{\mathcal{T}}_{\hat{X}_0\hat{\Xi}(\bullet)}\left[\frac{\partial\mathcal{G}[\upsilon\,;u(\bullet|_{t_0}^{t})]}{\partial\upsilon}\right].$$

Proof of the lemma for $\dfrac{\delta}{\delta u(s)}\left[\bar{\mathcal{T}}_{\hat{X}_0\hat{\Xi}(\cdot)}\,\mathcal{G}[\upsilon\,;u(\bullet|_{t_0}^{t})]\right]$ is similar, as it is also for the other two operators $\bar{\mathcal{T}}_{\hat{X}_0\hat{X}_0}$, $\bar{\mathcal{T}}_{\hat{\Xi}(\cdot)\hat{\Xi}(\cdot)}$. ■

**Proof of Lemma 3: Operators $\bar{\mathcal{T}}$ commute with each other.** For the sake of simplicity, we shall prove the commutativity of operators $\bar{\mathcal{T}}_{\hat{X}_0\hat{\Xi}(\cdot)}$ and $\bar{\mathcal{T}}_{\hat{X}_0\hat{X}_0}$. We begin from

$$\bar{\mathcal{T}}_{\hat{X}_0\hat{\Xi}(\cdot)}\,\bar{\mathcal{T}}_{\hat{X}_0\hat{X}_0}\,\mathcal{G}[\upsilon\,;u(\bullet|_{t_0}^{t})] =$$

by using Eqs. (B1a,b)

$$= \sum_{p=0}^{\infty}\frac{1}{p!}\int_{t_0}^{t}\overset{(p)}{\cdots}\int_{t_0}^{t} C_{X_0\Xi(\cdot)}(s^{(1)})\cdots C_{X_0\Xi(\cdot)}(s^{(p)})\times$$

$$\times\frac{\partial^p\,\delta^p}{\partial\upsilon^p\,\delta u(s^{(1)})\cdots\delta u(s^{(p)})}\left[\sum_{m=0}^{\infty}\frac{1}{m!}\frac{1}{2^m}C^m_{X_0X_0}\frac{\partial^{2m}\mathcal{G}[\upsilon\,;u(\bullet|_{t_0}^{t})]}{\partial\upsilon^{2m}}\right]ds^{(1)}\cdots ds^{(p)} =$$

rearranging the order of summations and using the linearity of derivatives and integrals

$$= \sum_{m=0}^{\infty}\frac{1}{m!}\frac{1}{2^m}C^m_{X_0X_0}\times$$

$$\times\frac{\partial^{2m}}{\partial\upsilon^{2m}}\left[\sum_{p=0}^{\infty}\frac{1}{p!}\int_{t_0}^{t}\overset{(p)}{\cdots}\int_{t_0}^{t} C_{X_0\Xi(\cdot)}(s^{(1)})\cdots C_{X_0\Xi(\cdot)}(s^{(p)})\,\frac{\partial^p\,\delta^p\,\mathcal{G}[\upsilon\,;u(\bullet|_{t_0}^{t})]}{\partial\upsilon^p\,\delta u(s^{(1)})\cdots\delta u(s^{(p)})}\,ds^{(1)}\cdots ds^{(p)}\right]$$

which, by Eqs. (B1a,b), is identified as

$$= \bar{\mathcal{T}}_{\hat{X}_0\hat{X}_0}\,\bar{\mathcal{T}}_{\hat{X}_0\hat{\Xi}(\cdot)}\,\mathcal{G}[\upsilon\,;u(\bullet|_{t_0}^{t})].$$ ■

In Lemmata 4-6, the actions of operators $\bar{\mathcal{T}}$ on the product $u(t)\,\mathcal{F}[\upsilon\,;u(\bullet)]$ are determined. For the actions of $\bar{\mathcal{T}}_{\hat{X}_0\hat{\Xi}(\cdot)}$ $\bar{\mathcal{T}}_{\hat{\Xi}(\cdot)\hat{\Xi}(\cdot)}$, the following *product rule for Volterra derivatives* is also needed

$$\frac{\delta^k\left[u(t)\,\mathcal{F}[\upsilon\,;u(\bullet|_{t_0}^{t})]\right]}{\delta u(s^{(1)})\cdots\delta u(s^{(k)})} = u(t)\,\frac{\delta^k\,\mathcal{F}[\upsilon\,;u(\bullet|_{t_0}^{t})]}{\delta u(s^{(1)})\cdots\delta u(s^{(k)})} + \sum_{n=1}^{k}\delta(t-s^{(n)})\,\frac{\delta^{k-1}\mathcal{F}[\upsilon\,;u(\bullet|_{t_0}^{t})]}{\prod_{\substack{\ell=1\\ \ell\neq n}}^{k}\delta u(s^{(\ell)})}, \tag{B2}$$

with $\prod_{\substack{\ell=1\\ \ell\neq n}}^{k}\delta u(s^{(\ell)}) = \delta u(s^{(1)})\cdots\delta u(s^{(n-1)})\,\delta u(s^{(n+1)})\cdots\delta u(s^{(k)})$. Eq. (B2) is easily proven via mathematical induction on index $k$, commencing from the relation for $k=1$ (product rule for the first-order derivative):

$$\frac{\delta\left[u(t)\,\mathcal{F}[\upsilon\,;u(\bullet|_{t_0}^{t})]\right]}{\delta u(s)} = u(t)\,\frac{\delta\mathcal{F}[\upsilon\,;u(\bullet|_{t_0}^{t})]}{\delta u(s)} + \frac{\delta u(t)}{\delta u(s)}\,\mathcal{F}[\upsilon\,;u(\bullet|_{t_0}^{t})],$$

in which $\delta u(t)/\delta u(s)=\delta(t-s)$. Having formulated Eq. (B2), we move on to the proofs of Lemmata 4-6.

**Proof of Lemma 4:** The action of operator $\bar{\mathcal{T}}_{\hat{X}_0\hat{X}_0}$ on $u(t)\,\mathcal{F}[\upsilon\,;u(\bullet|_{t_0}^{t})]$ is given by

$$\bar{\mathcal{T}}_{\hat{X}_0\hat{X}_0}\left[u(t)\,\mathcal{F}[\upsilon\,;u(\bullet|_{t_0}^{t})]\right] = u(t)\,\bar{\mathcal{T}}_{\hat{X}_0\hat{X}_0}\left[\mathcal{F}[\upsilon\,;u(\bullet|_{t_0}^{t})]\right]. \tag{B3}$$

Eq. (B3) holds true due to the linearity of $\bar{\mathcal{T}}_{\hat{X}_0\hat{X}_0}$, since the factor $u(t)$ is independent from $\upsilon$, which is the differentiation argument of $\bar{\mathcal{T}}_{\hat{X}_0\hat{X}_0}$. ■

**Proof of Lemma 5.** By using the series expansion of Eq. (B1b), and employing the product rule, Eq. (B2), the action of operator $\bar{\mathcal{T}}_{\hat{X}_0\hat{\Xi}(\bullet)}$ on $u(t)\;\mathcal{F}[\upsilon\,;u(\bullet|_{t_0}^{t})]$ is expressed as

$$\bar{\mathcal{T}}_{\hat{X}_0\hat{\Xi}(\bullet)}\left[u(t)\;\mathcal{F}[\upsilon\,;u(\bullet|_{t_0}^{t})]\right] = u(t)\;\mathrm{A} + C_{X_0\Xi(\bullet)}(t)\;\mathrm{B}, \tag{B4}$$

where $\mathrm{A}$, $\mathrm{B}$ are

$$\mathrm{A} = \sum_{p=0}^{\infty}\frac{1}{p!}\int_{t_0}^{t}\overset{(p)}{\cdots}\int_{t_0}^{t} C_{X_0\Xi(\bullet)}(s^{(1)})\cdots C_{X_0\Xi(\bullet)}(s^{(p)})\,\frac{\partial^p\delta^p\mathcal{F}[\upsilon\,;u(\bullet|_{t_0}^{t})]}{\partial\upsilon^p\,\delta u(s^{(1)})\cdots\delta u(s^{(p)})}\,ds^{(1)}\cdots ds^{(p)},$$

$$\mathrm{B} = \sum_{p=1}^{\infty}\frac{1}{p!}\sum_{n=1}^{p}\int_{t_0}^{t}\overset{(p-1)}{\cdots}\int_{t_0}^{t}\prod_{\substack{m=1\\ m\neq n}}^{p} C_{X_0\Xi(\bullet)}(s^{(m)})\,\frac{\partial^p\delta^{p-1}\mathcal{F}[\upsilon\,;u(\bullet|_{t_0}^{t})]}{\partial\upsilon^p\prod_{\substack{m=1\\ m\neq n}}^{p}\delta u(s^{(\ell)})}\prod_{\substack{m=1\\ m\neq n}}^{p} ds^{(m)}.$$

By the symmetry of integration arguments, all terms of the $n-$sum are equal, thus,

$$\mathrm{B} = \sum_{p=1}^{\infty}\frac{1}{(p-1)!}\int_{t_0}^{t}\overset{(p-1)}{\cdots}\int_{t_0}^{t} C_{X_0\Xi(\bullet)}(s^{(1)})\cdots C_{X_0\Xi(\bullet)}(s^{(p-1)})\,\frac{\partial^p\delta^{p-1}\mathcal{F}[\upsilon\,;u(\bullet|_{t_0}^{t})]}{\partial\upsilon^p\,\delta u(s^{(1)})\cdots\delta u(s^{(p-1)})}\,ds^{(1)}\cdots ds^{(p-1)}.$$

In view of Eq. (B1b), series $\mathrm{A}$ is identified as

$$\mathrm{A} = \bar{\mathcal{T}}_{\hat{X}_0\hat{\Xi}(\bullet)}\left[\mathcal{F}[\upsilon\,;u(\bullet|_{t_0}^{t})]\right]. \tag{B5}$$

For series $\mathrm{B}$, we perform the index change $k=p-1$, resulting in

$$\mathrm{B} = \sum_{k=0}^{\infty}\frac{1}{k!}\int_{t_0}^{t}\overset{(k)}{\cdots}\int_{t_0}^{t} C_{X_0\Xi(\bullet)}(s^{(1)})\cdots C_{X_0\Xi(\bullet)}(s^{(k)})\,\frac{\partial^k\delta^k}{\partial\upsilon^k\,\delta u(s^{(1)})\cdots\delta u(s^{(k)})}\left[\frac{\partial\mathcal{F}[\upsilon\,;u(\bullet|_{t_0}^{t})]}{\partial\upsilon}\right]ds^{(1)}\cdots ds^{(k)} \tag{B6}$$

Via Eq. (B1b), the right-hand side of Eq. (B6) is identified as

$$\mathrm{B} = \bar{\mathcal{T}}_{\hat{X}_0 \hat{\Xi}(\bullet)} \left[ \frac{\partial \mathcal{F}[\upsilon ; u(\bullet|_{t_0}^{t})]}{\partial \upsilon} \right]. \tag{B7}$$

Substitution of Eqs. (B5), (B7) into Eq. (B4) results in Eq. (25) of the main paper for $\mathcal{F}[\upsilon ; u(\bullet|_{t_0}^{t})] = \mathcal{F}_1[\upsilon ; u(\bullet|_{t_0}^{t})]$, completing thus the proof of Lemma 5 ■

**Proof of Lemma 6.** By expanding $\bar{\mathcal{T}}_{\hat{\Xi}(\bullet)\hat{\Xi}(\bullet)} \left[ u(t)\, \mathcal{F}[\upsilon ; u(\bullet|_{t_0}^{t})] \right]$ in series using Eq. (B1c), and employing the product rule, Eq. (B2), we obtain

$$\bar{\mathcal{T}}_{\hat{\Xi}(\bullet)\hat{\Xi}(\bullet)} \left[ u(t)\, \mathcal{F}[\upsilon ; u(\bullet|_{t_0}^{t})] \right] = u(t)\,\Gamma + \Delta, \tag{B8}$$

where $\Gamma$ is

$$\Gamma = \sum_{p=0}^{\infty} \frac{1}{p!} \frac{1}{2^p} \int_{t_0}^{t} \overset{(2p)}{\cdots} \int_{t_0}^{t} \Big[ C_{\Xi(\bullet)\Xi(\bullet)}(s_1^{(1)}, s_2^{(1)}) \cdots C_{\Xi(\bullet)\Xi(\bullet)}(s_1^{(p)}, s_2^{(p)}) \times \\ \times \frac{\delta^{2p} \mathcal{F}[\upsilon ; u(\bullet|_{t_0}^{t})]}{\delta u(s_1^{(1)}) \delta u(s_2^{(1)}) \cdots \delta u(s_1^{(p)}) \delta u(s_2^{(p)})} \, ds_1^{(1)} ds_2^{(1)} \cdots ds_1^{(p)} ds_2^{(p)} \Big],$$

while $\Delta$ can be written as follows, after taking into account the symmetry of autocorrelation function $C_{\Xi(\bullet)\Xi(\bullet)}(s_1, s_2)$:

$$\Delta = \sum_{p=1}^{\infty} \frac{1}{p!} \frac{1}{2^p} \sum_{n=1}^{2p} \int_{t_0}^{t} \overset{(2p-1)}{\cdots} \int_{t_0}^{t} \Big[ C_{\Xi(\bullet)\Xi(\bullet)}(t, s_2^{(n)}) \prod_{\substack{m=1 \\ m \neq n}}^{p} C_{\Xi(\bullet)\Xi(\bullet)}(s_1^{(m)}, s_2^{(m)}) \times \\ \times \frac{\delta^{2p-1} \mathcal{F}[\upsilon ; u(\bullet|_{t_0}^{t})]}{\delta u(s_2^{(n)}) \prod_{\substack{m=1 \\ m \neq n}}^{p} \delta u(s_1^{(m)}) \delta u(s_2^{(m)})} \, ds_2^{(n)} \prod_{\substack{m=1 \\ m \neq n}}^{p} ds_1^{(m)} ds_2^{(m)} \Big].$$

By virtue of Eq. (B1c), series $\Gamma$ is identified as

$$\Gamma = \bar{\mathcal{T}}_{\hat{\Xi}(\bullet)\hat{\Xi}(\bullet)} \left[ \mathcal{F}[\upsilon ; u(\bullet|_{t_0}^{t})] \right]. \tag{B9}$$

We turn now our attention to the series $\Delta$. By performing, in each term of the $n-$sum $\sum_{n=1}^{2p} \cdots$, the change of integration variables:

$s = s_2^{(n)}$ and $s_i^{(\ell)} = s_i^{(m)}$ for $m < n$, $s_i^{(\ell)} = s_i^{(m-1)}$ for $m > n$, $i = 1, 2$, we obtain:

$$\prod_{\ell=1}^{p-1} C_{\Xi(\bullet)\Xi(\bullet)}(s_1^{(\ell)}, s_2^{(\ell)}) = \prod_{\substack{m=1 \\ m \neq n}}^{p} C_{\Xi(\bullet)\Xi(\bullet)}(s_1^{(m)}, s_2^{(m)}),$$

$$\prod_{\ell=1}^{p-1} \delta u(s_1^{(\ell)}) \delta u(s_2^{(\ell)}) = \prod_{\substack{m=1 \\ m \neq n}}^{p} \delta u(s_1^{(m)}) \delta u(s_2^{(m)}),$$

$$\prod_{\ell=1}^{p-1} ds_1^{(\ell)}\, ds_2^{(\ell)} = \prod_{\substack{m=1 \\ m \neq n}}^{p} ds_1^{(m)}\, ds_2^{(m)}.$$

Now, it is easy to see that all $2p$ terms in $n-$sum are equal, resulting in

$$\Delta = \sum_{p=1}^{\infty} \frac{1}{(p-1)!} \frac{1}{2^{p-1}} \int_{t_0}^{t} \overset{(2p-1)}{\cdots} \int_{t_0}^{t} \Big[ C_{\Xi(\bullet)\Xi(\bullet)}(t,s)\, C_{\Xi(\bullet)\Xi(\bullet)}(s_1^{(1)}, s_2^{(1)}) \cdots C_{\Xi(\bullet)\Xi(\bullet)}(s_1^{(p-1)}, s_2^{(p-1)}) \times$$

$$\times \frac{\delta^{2p-1} \mathcal{F}[\upsilon\,; u(\bullet|_{t_0}^{t})]}{\delta u(s) \delta u(s_1^{(1)}) \delta u(s_2^{(1)}) \cdots \delta u(s_1^{(p-1)}) \delta u(s_2^{(p-1)})} ds\, ds_1^{(1)}\, ds_2^{(1)} \cdots ds_1^{(p-1)}\, ds_2^{(p-1)} \Bigg].$$

Further, we perform the index change $k = p-1$ and interchange $s-$integration with summation, resulting in

$$\Delta = \int_{t_0}^{t} C_{\Xi(\bullet)\Xi(\bullet)}(t,s) \sum_{k=0}^{\infty} \frac{1}{k!} \frac{1}{2^k} \int_{t_0}^{t} \overset{(2k)}{\cdots} \int_{t_0}^{t} \Big[ C_{\Xi(\bullet)\Xi(\bullet)}(s_1^{(1)}, s_2^{(1)}) \cdots C_{\Xi(\bullet)\Xi(\bullet)}(s_1^{(k)}, s_2^{(k)}) \times$$

$$\times \frac{\delta^{2k}}{\delta u(s_1^{(1)}) \delta u(s_2^{(1)}) \cdots \delta u(s_1^{(k)}) \delta u(s_2^{(k)})} \left[ \frac{\delta \mathcal{F}[\upsilon\,; u(\bullet|_{t_0}^{t})]}{\delta u(s)} \right] ds_1^{(1)}\, ds_2^{(1)} \cdots ds_1^{(k)}\, ds_2^{(k)} \Bigg] ds. \tag{B10}$$

The sum in the right-hand side of Eq. (B10) is identified, via Eq. (B1c), as

$$\Delta = \int_{t_0}^{t} C_{\Xi(\bullet)\Xi(\bullet)}(t,s)\, \bar{\mathcal{T}}_{\hat{\Xi}(\bullet)\hat{\Xi}(\bullet)} \left[ \frac{\delta \mathcal{F}[\upsilon\,; u(\bullet|_{t_0}^{t})]}{\delta u(s)} \right] ds. \tag{B11}$$

By substituting Eqs. (B9), (B11) into Eq. (B8), we obtain Eq. (28) of the main paper for $\mathcal{F}[\upsilon\,; u(\bullet|_{t_0}^{t})] = \mathcal{F}_2[\upsilon\,; u(\bullet|_{t_0}^{t})]$. Thus, proof of Lemma 6 is completed. ■